\documentclass[12pt,preprint]{aastex}

\newcommand{\ez}{\mbox{$\hat{{\bf e}}_{\rm z}$}}

\shorttitle{Damped oscillations in two inhomogeneous coronal slabs}
\shortauthors{Arregui et al.}

\begin{document}

\title{The resonant damping of fast magnetohydrodynamic oscillations in a system of two coronal slabs}

\author{I\~nigo Arregui\altaffilmark{1}, Jaume
Terradas\altaffilmark{1,2}, Ram\'on Oliver\altaffilmark{1}, and
Jos\'e Luis Ballester\altaffilmark{1}}


\altaffiltext{1}{Departament de F\'{\i}sica, Universitat de les Illes Balears,
E-07122 Palma de Mallorca, Spain. Email: inigo.arregui@uib.es,
jaume.terradas@uib.es, ramon.oliver@uib.es and dfsjlb0@uib.es}
\altaffiltext{2}{Centrum voor Plasma Astrofysica, K.U. Leuven,
Celestijnenlaan 200B, B-3001 Heverlee, Belgium}

\begin{abstract}
Observations of transversal coronal loop oscillations very often show the excitation and damping
of oscillations in groups of coronal loops rather than in individual and isolated structures.
We present results on the oscillatory properties (periods, damping rates, and spatial distribution of 
perturbations) for resonantly damped oscillations in a system of two inhomogeneous coronal slabs and compare
them to the  properties found in single slab loop models.
A system of two identical coronal loops is modeled, in Cartesian geometry, as being composed by 
two density enhancements. The linear magnetohydrodynamic (MHD) wave equations for oblique propagation of waves are 
solved and the damping of the different solutions, due to the transversal inhomogeneity of the density profile, is computed. 
The physics of the obtained results is analyzed by an examination of the perturbed physical variables.
We find that, due to the interaction between the loops, the normal modes of oscillation present in a single slab  split 
into symmetric and antisymmetric oscillations when a system of two identical slabs is considered. 
The frequencies of these solutions may differ from  the single slab results when the distance between the loops is of the 
order of a few slab widths. Oblique propagation of waves weakens this interaction, since 
solutions become more confined to the edges of the slabs.  The damping is strong for 
surface-like oscillations, while sausage body-like solutions are unaffected. For some solutions, and small 
slab separations, the damping in a system of two loops differs substantially from the damping of a single loop.
\end{abstract}


\section{INTRODUCTION}\label{intro}


The solar corona, the outermost part of the solar atmosphere, is able to support a 
variety of MHD waves in its different magnetic and plasma configurations. 
In the last years, particular attention has been devoted to the phenomenon of 
transversal coronal loop oscillations, first observed by instruments
on-board TRACE spacecraft \citep{Aschwanden99,Nakariakov99,Aschwanden02, SAT02}. These oscillations, 
triggered mainly by nearby flares or filament eruptions, consist of lateral displacements
of the loops while their foot-points are fixed to the denser photosphere. The characteristic
periods are of the order of minutes and oscillations are quickly damped in a few periods.

From a theoretical point of view, these oscillations  have been interpreted by \citet{Nakariakov99}
as the fundamental fast MHD kink mode of a flux tube. This 
solution involves a global motion of the whole structure that displaces the axis of the tube.
The identification of observed oscillations  with theoretical MHD wave solutions
is the key for coronal seismology,  first suggested by \citet{uchida70,REB84}, and has 
allowed preliminary estimations of the magnetic field strength, transport coefficients, and 
plasma-$\beta$ in coronal loops \citep{Nakariakov99,Nakariakov01,markusbook}. Some recent examples of the application 
of coronal seismology can be found in \cite{Nakariakov01,GAA02,AAG05,verwichte06,Arregui07}. 
As for the nature of the damping mechanism(s) there is little consensus yet and 
several mechanisms are currently under study, such as  non-ideal effects, lateral wave 
leakage due to the curvature of the loops \citep{TOB06}, mechanisms based on the 
topology of the magnetic field lines \citep{SB00}, and resonant conversion of wave energy, due 
to the non-uniformity of the magnetic-plasma configuration 
\citep{HY88,RR02,GAA02,GAA06}. Foot-point leakage through the chromospheric 
density gradient \citep{depontieu01,ofman02}  has been invoked for the 
damping of Alfv\'en waves and could be ascribed to the damping of transverse 
oscillations, as well.


Observations of transversal coronal loop oscillations very often show the excitation 
and damping of motions in groups of coronal loops rather than in single, isolated 
structures (see for example the event recorded by TRACE on 2001 April 15,  
analyzed by \citealt{Verwichte04}). However, most of the theoretical models used for their 
study are based on single loop models. The interaction between the 
motions of adjacent coronal loops may change the oscillatory properties (periods, damping
times, and spatial distribution of perturbations) of the observed event. For this reason, 
it is important to fully understand the theoretical properties of MHD waves in 
multiple-loop structures. The interest of these studies is two-fold. On one hand, to see under which
conditions these differences in the oscillatory properties are important and, on the other hand, if important, 
to make reliable predictions and determinations of unknown physical  parameters in the corona. 
For example, when analyzing the 2001 April 15 event, \citet{Verwichte04} report on damping times that are longer 
than expected from previous observations of transverse oscillations in isolated loops. 
The fact that the studied coronal loops belong to a group could be behind this 
result.


Several authors have previously investigated the properties of MHD waves in multi-structures.
\citet{BH87} studied the properties of the normal modes of oscillation of a periodic medium. 
\citet{BF91} and \citet{keppens94} considered the scattering and absorption of acoustic waves 
by bundles of magnetic flux tubes with sunspot properties. \citet{murawski93} and \citet{MR94}
studied, by means of numerical simulations, the propagation of fast waves in a system of two slabs 
unbounded in the longitudinal direction. The collective nature and properties of oscillations in multi-fibril 
Cartesian systems was analyzed by \citet{diaz05}, in the context of prominence oscillations.
More recently, \citet{ofman05} performed numerical simulations for the damping of a bundle of four loops, 
\citet{marcu06} have extended the study by \citet{BH87} by including a steady motion in the equilibrium and
\citet{OM07} have performed numerical simulations of impulsively excited magneto-sonic waves in two 
parallel solar coronal slabs.
Also recently, \citet{Luna06} have explored the normal modes and the time-dependent 
evolution of fast kink MHD perturbations in a system of two coronal loops, modeled as plasma slabs.  
They find a splitting of modes into symmetric and antisymmetric kink-like oscillations.
The interaction between two coronal loops produces a variation on the frequency and spatial
structure of the eigenfunctions, that depend now on the distance between
the loops. 


In this paper, we present the oscillatory properties of fast MHD waves in a system of two Cartesian slabs when both oblique 
propagation of perturbations and transversal inhomogeneity of 
the medium are included. These two ingredients produce the resonant damping of fast waves.
A particular study of the oscillatory properties of a similar physical  system  was 
recently presented by \citet{ATOB07a}, who considered two very close slabs to analyze the effect of the possibly 
unresolved internal structuring of coronal loops on their transverse damped oscillations.
Here a more in-depth analysis is performed. The resonant damping of fast MHD oscillations in a single Cartesian slab has been 
described by \citet{ATOB07b}. In this paper, a second density enhancement is included and an analogous analysis is 
presented. This work, thus, constitutes an extension to \citet{Luna06} to the case of oblique propagation and 
damping by resonant absorption and an extension to \citet{ATOB07b} to a system of two interacting 
slabs. Although the model adopted in this work is an oversimplification, and is far from being an accurate
representation of real coronal loop systems, this simplicity allows a detailed study of the problem.


The layout of the paper is as follows. In  \S~\ref{equil}, the equilibrium
configuration, the  linear MHD wave equations for coupled fast and Alfv\'en waves, and a description
of the numerical method are presented. Then, \S~\ref{results} describes the results of our study.
First, in \S~\ref{undamped}, the normal mode properties of undamped fast waves for oblique propagation 
are described. Then, in \S~\ref{damping}, the damping properties of the solutions and the 
physical interpretation of the  results are discussed.  Finally,
in \S~\ref{conclusions}, our 
conclusions are drawn.

\section{EQUILIBRIUM MODEL, LINEAR MHD WAVE EQUATIONS, AND NUMERICAL METHOD}\label{equil}

\begin{figure}
\centering
\includegraphics[width=10.0cm,angle=90]{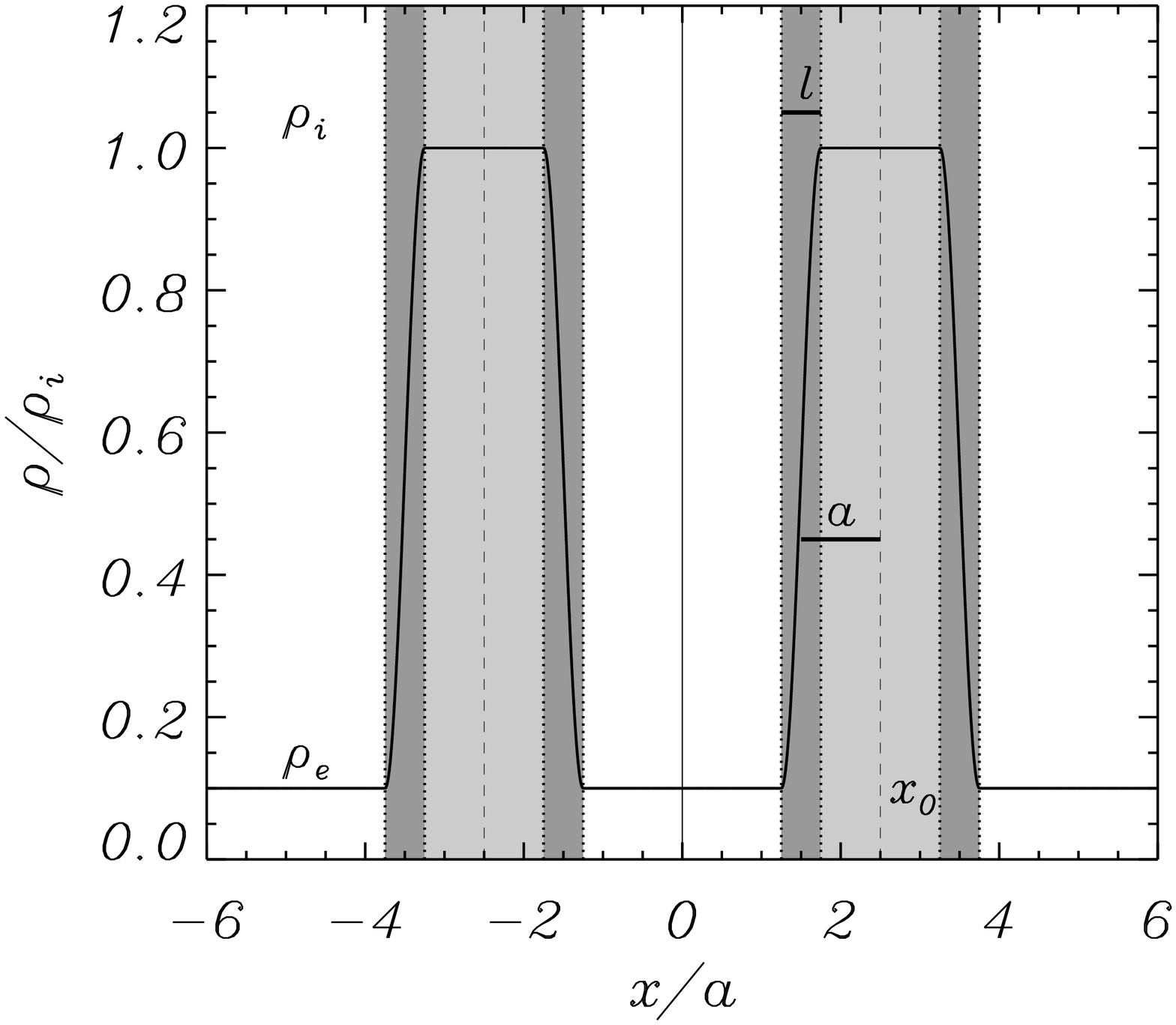}
\caption{Schematic representation of the two density enhancements
      (light-shaded regions) of half-width $a$, centered at
      $\pm x_0$ representing a system of two coronal slabs in the direction transverse to 
      the equilibrium magnetic field. These enhancements with internal density $\rho_i$ connect 
      to the external medium, with density $\rho_e$, by
      transitional non-uniform layers 
      (shaded regions) of
      thickness $l$.}
\label{model}
\end{figure}

We model the equilibrium magnetic and plasma
configuration of a system of two coronal loops by means
of a one-dimensional model in Cartesian geometry. 
The magnetic field is straight and pointing in the $z$-direction, ${\mathit {\bf  B}}=B\ {\bf \ez}$. 
For applications to the solar corona, it is a good approximation to consider that
the magnetic pressure dominates over the gas
pressure. This classic zero plasma-$\beta$ limit 
implies that the magnetic field is uniform and
that the density, $\rho(x)$, or Alfv\'en
speed, $v_{\rm A}(x)$, profiles can be chosen arbitrarily. 
For simplicity, we consider a system of two identical coronal loops.
This system is then modeled by defining a particular equilibrium density
profile in the $x$-direction (see Figure~\ref{model}), with two density enhancements 
of half-width $a$ located at $\pm x_0$.  The density in each of the slabs is uniform, $\rho_i$, and
connected to the uniform coronal environment, with density $\rho_e$, by transitional
non-uniform layers of thickness $l$. The explicit expression for the considered
equilibrium density is (for $x \ge 0$)

\begin{math}
\rho(x) =
   \cases{
        \rho_e                & if \  $0\leq x \leq x_0-a-\frac{l}{2}$,\cr
         f_1(x) 	      & if \ $x_0-a-\frac{l}{2}\leq x \leq x_0-a+\frac{l}{2}$,\cr
         \rho_i               & if \ $x_0-a-\frac{l}{2}\leq x \leq x_0-a+\frac{l}{2}$,\cr
          f_2(x)              & if \ $x_0+a-\frac{l}{2}\leq x \leq x_0+a+\frac{l}{2}$,\cr
	  \rho_e              & if \ $x\geq  x_0+a+\frac{l}{2}$,\cr
     }
\end{math}

\noindent
with $\rho(-x)=\rho(x)$.
The particular dependence of the density profiles at the non-uniform transitional layers
has no qualitative importance on the obtained results. We have chosen them, following \cite{RR02} 
and \citet{tom04b}, as

\begin{eqnarray}
f_1(x)&=&\frac{\rho_i}{2}\left[\left(1+\frac{\rho_e}{\rho_i}\right)-\left(1-\frac{\rho_e}{\rho_i}
\right)\sin{\frac{\pi\left[-x-(-x_0+a)\right]}{l}}\right]\\\nonumber
\end{eqnarray}

\noindent
and

\begin{eqnarray}
f_2(x)&=&\frac{\rho_i}{2}\left[\left(1+\frac{\rho_e}{\rho_i}\right)-\left(1-\frac{\rho_e}{\rho_i}
\right)\sin{\frac{\pi\left[x-(x_0+a)\right]}{l}}\right].\label{f2}
\end{eqnarray}

This equilibrium configuration is the same as the one used  by
\citet{ATOB07a} for the study of the influence of the internal density structuring of a coronal loop
on the properties of their damped oscillations. Their analysis is based on an equilibrium model 
in which the two slabs are very close to each other and contains a comparison of the results to the ones obtained 
with similar single slab models. This equilibrium model is also an extension of the one considered 
by \citet{Luna06}, with the addition of  non-uniform density layers at the edges of each slab. 


In order to study small amplitude oscillations of the previous equilibrium,
the linear resistive MHD equations, with constant magnetic diffusivity, $\eta$, are considered. 
As the equilibrium configuration only
depends on the $x$-direction, a spatial and temporal dependence of the form
$\exp^{\imath(\omega t-k_y y - k_z z)}$ is assumed for all perturbed quantities,
with $\omega=\omega_R+\imath\omega_I$ the complex frequency
and $k_y$ and $k_z$ the perpendicular and parallel
wavenumbers. For resonantly damped solutions, the real 
part of the frequency gives the period of the oscillation, $P=2\pi/\omega_R$, while the imaginary part 
is related to the damping time, $\tau_d=1/\omega_I$.  
The photospheric line-tying effect is then
included by selecting the appropriate parallel
wavenumber. 
This leads to the following set of ordinary differential equations for the two components of the 
velocity perturbation, $v_x$ and $v_y$, and the three components of the perturbed magnetic field, $b_x$,
$b_y$, and $b_z$,

\begin{eqnarray}
\omega v_x &=& \frac{B}{\rho}\left(- k_z b_{\rm 1x}+\imath \frac{d b_{\rm 1z}}{dx}\right), \label{first}\\
\omega v_y &=& \frac{B}{\rho}\left(- k_z b_{\rm 1y}+ k_y b_{\rm 1z}\right),\\
\omega \frac{b_{\rm 1x}}{B}&=&- k_z v_x -\imath\frac{\eta}{B}\left[\frac{d^2 b_{\rm 1x}}{dx^2}-\left(k^2_y+k^2_z\right) b_{\rm 1x}\right],\\
\omega \frac{b_{\rm 1y}}{B}&=&- k_z v_y -\imath\frac{\eta}{B}\left[\frac{d^2 b_{\rm 1y}}{dx^2}-\left(k^2_y+k^2_z\right) b_{\rm 1y}\right],\\
\omega \frac{b_{\rm 1z}}{B}&=&\left(\imath\frac{dv_x}{dx}+ k_y v_y\right)-\imath\frac{\eta}{B}
\left[\frac{d^2 b_{\rm 1z}}{dx^2}-\left(k^2_y+k^2_z\right) b_{\rm 1z}\right].\label{last}
\end{eqnarray}

\noindent
These equations, together with the appropriate boundary conditions, define an eigenvalue problem 
for fast and Alfv\'en normal modes.
As the plasma-$\beta$=$0$, the slow mode is absent and there 
are no motions parallel to the equilibrium magnetic field, $v_z=0$.
In this paper, solutions to these equations are obtained by performing a normal mode analysis.
When only perpendicular propagation of perturbations is considered ($k_y\neq 0$, $l=0$), the solutions 
represent undamped fast MHD waves with oblique propagation, in addition to Alfv\'en waves.
If non-uniform transitional layers are included ($l\neq 0$), the 
solutions are damped. In this case, solutions are in general difficult to obtain analytically. 
Numerical approximations  are then obtained 
using PDE2D \citep{sewell05}, a general-purpose partial differential equation solver. The code uses finite elements
and allows the use of a non-uniformly distributed grid, which is needed in order to properly resolve the large gradients
that arise in the vicinity of resonant layers. This code  has been used successfully in  
previous studies that involved similar computations of the damping rate by, for example, 
\citet{TOB06,ATOB07a,ATOB07b}.
As for the boundary conditions, we impose the vanishing of the perturbed 
velocity far away from the two-slab system, hence ${\bf v}\rightarrow 0$ as $x\rightarrow\pm\infty$.

\section{ANALYSIS AND RESULTS}\label{results}

Solutions to Equations~(\ref{first})--(\ref{last}) for the simplest case, 
when $k_y=0$ and $l=0$, have been obtained by \citet{Luna06}. 
Fast normal modes are then decoupled from Alfv\'en modes. \citet{Luna06} find that the system is able 
to support a variety of oscillatory modes. In particular, they concentrate on the splitting of the kink 
transversal oscillation of a single slab, giving rise to two solutions, symmetric and 
antisymmetric  with respect to $x=0$, in  which the two slabs oscillate in the kink mode in 
phase or anti-phase 
with the same frequency. The fundamental mode is symmetric and its
frequency is constrained between two limiting values. In the limit of no separation between slabs, it is equal to the  
kink mode frequency of a single slab with double width. As the distance between the slabs is 
increased, the interaction weakens and, for large separations, the two slabs do not ``feel'' each 
other and the frequency of the whole system is the same as the kink mode frequency of a single slab.
The range of separation between slabs for which
the interaction is strong is mainly determined by
the parallel wavenumber, 
$k_z$, which determines the confinement and drop-off rate of the eigenfunctions in the coronal medium. 
These authors also find that the  
antisymmetric mode is not present for all the values of the distance 
between the slabs and becomes leaky when this distance is smaller than a given value, that 
depends on the adopted value for the parallel
wavenumber.

When compared to \citet{Luna06}, there are two new ingredients in the physical model under 
consideration in this paper. On one hand, the non-uniformity of the equilibrium density at the 
edges of the slabs. 
On the other hand, the inclusion of perpendicular propagation of perturbations. 
These two ingredients produce the resonant coupling of fast modes to Alfv\'en modes and, as a 
result, the global modes of the system are damped. The damping of oscillations in a system of 
coronal loops has been reported in observations analyzed by \citet{Verwichte04} and is likely to differ 
from the theoretical values obtained with isolated coronal loop models. 

In \S~\ref{undamped}, we first describe 
the normal mode properties of oscillations of our two-slab system when oblique propagation of  perturbations
is included. Then, in \S~\ref{damping}, the damping of oscillations by resonant absorption and the properties of 
the eigenfunctions are detailed.

\subsection{Normal Modes in a Two-Slab System with Oblique Propagation}\label{undamped}

We first consider solutions to Equations~(\ref{first})--(\ref{last}) including oblique propagation of 
waves ($k_y\neq0$) and in the absence of  non-uniform transitional layers ($l=0$).
Also, $\eta=0$ is set in these equations.
 As the equilibrium is piece-wise constant there is no resonant coupling to 
Alfv\'en waves and solutions represent undamped oscillations. These solutions can be obtained by solving an analytical dispersion
relation. \citet{ATOB07b} have shown,  for a single Cartesian slab, that oblique propagation of waves affects 
the nature and properties of kink- and sausage-like solutions. The fundamental kink mode, which is body for
values of $k_y$ below certain value, becomes surface beyond that particular value.
There is a sausage surface wave, with its frequency always below the internal cut-off frequency, which 
only exists when $k_y\neq0$. The phase speed of these two solutions goes asymptotically to the kink speed for 
quasi-perpendicular propagation ($k_y\gg k_z$). On the other hand, the sausage body solution, which is leaky in the 
long wave limit ($k_z a\ll 1$), becomes trapped when a non-zero $k_y$ is included. This solution is always above the 
internal cut-off frequency and, hence, keeps its body character in the limit of quasi-perpendicular 
propagation, due to the existence of the sausage surface solution below the internal cut-off frequency.
All these features are well displayed in Figure~2 of \citet{ATOB07b}.

When a second density enhancement is included, the kink, sausage surface, and 
sausage body solutions, described by \citet{ATOB07b}, split into symmetric and antisymmetric 
solutions with respect to $x=0$, giving six solutions that we  name symmetric and antisymmetric; 
kink (with the two slabs oscillating  in the kink mode), sausage surface (with the two 
slabs oscillating in the sausage surface mode), and sausage body (with both slabs oscillating in 
the sausage body mode).

\begin{figure}
   \includegraphics[width=7.0cm,angle=0]{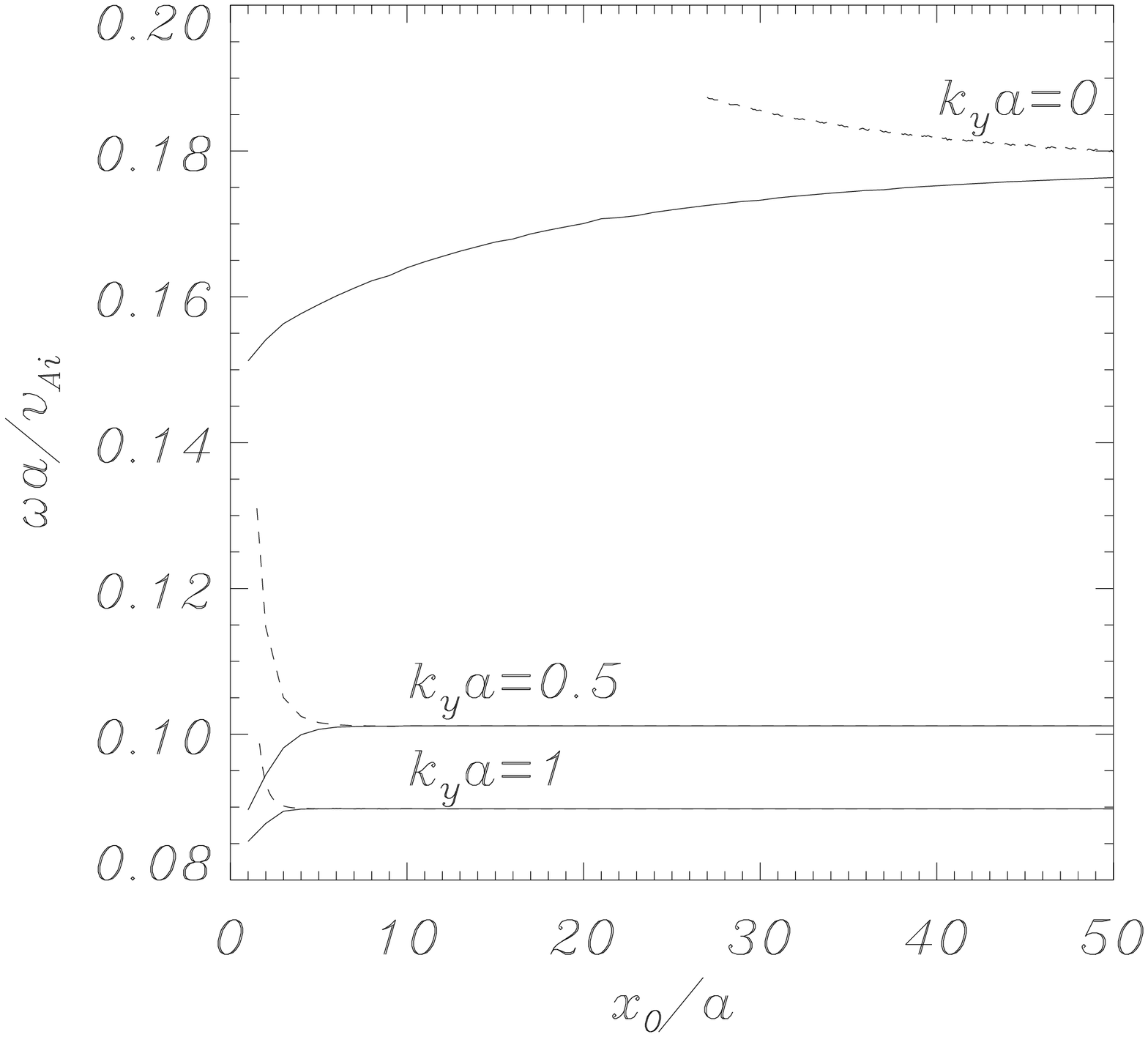}
  \includegraphics[width=7.0cm,angle=0]{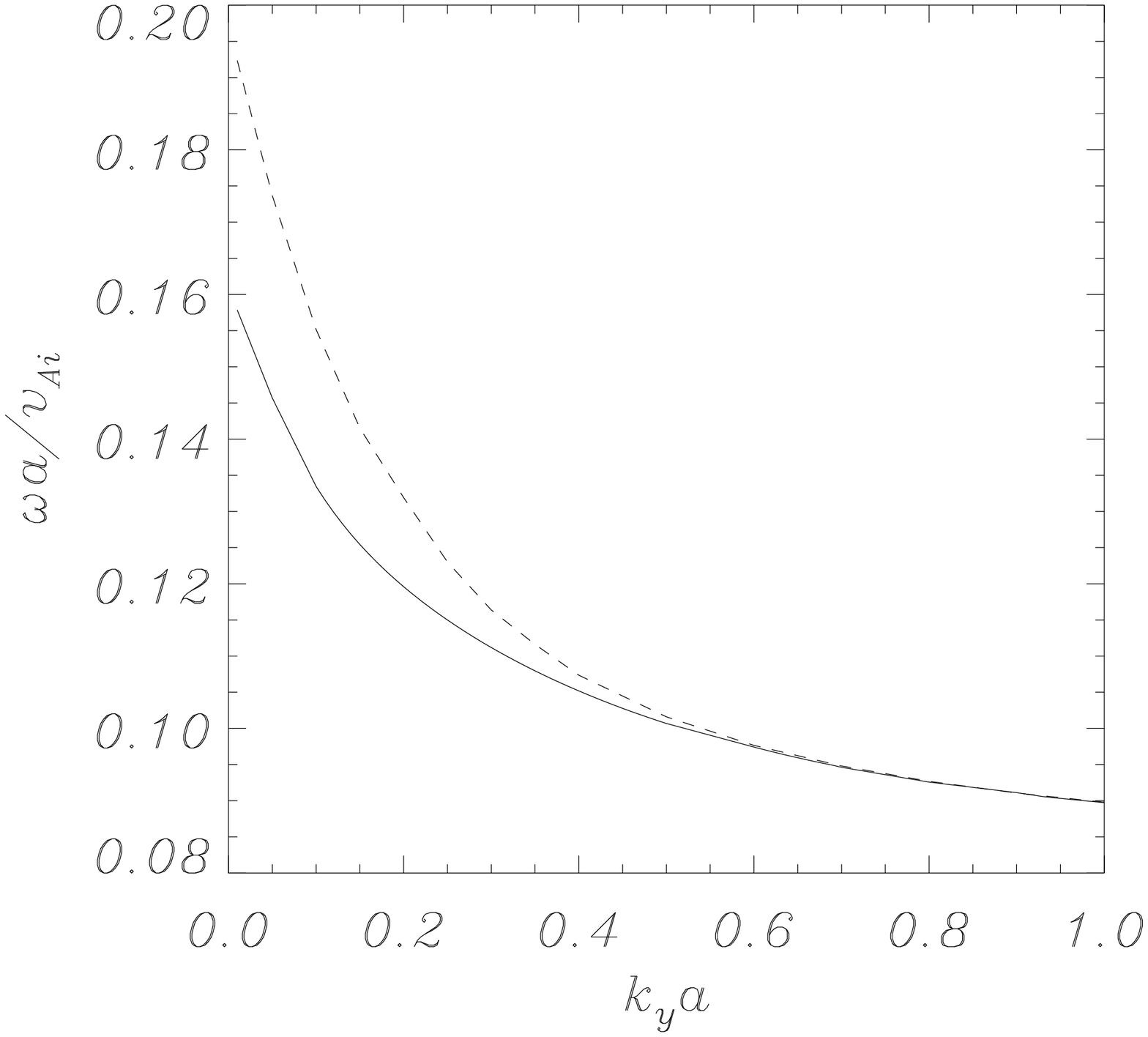}\\
   \includegraphics[width=7.0cm,angle=0]{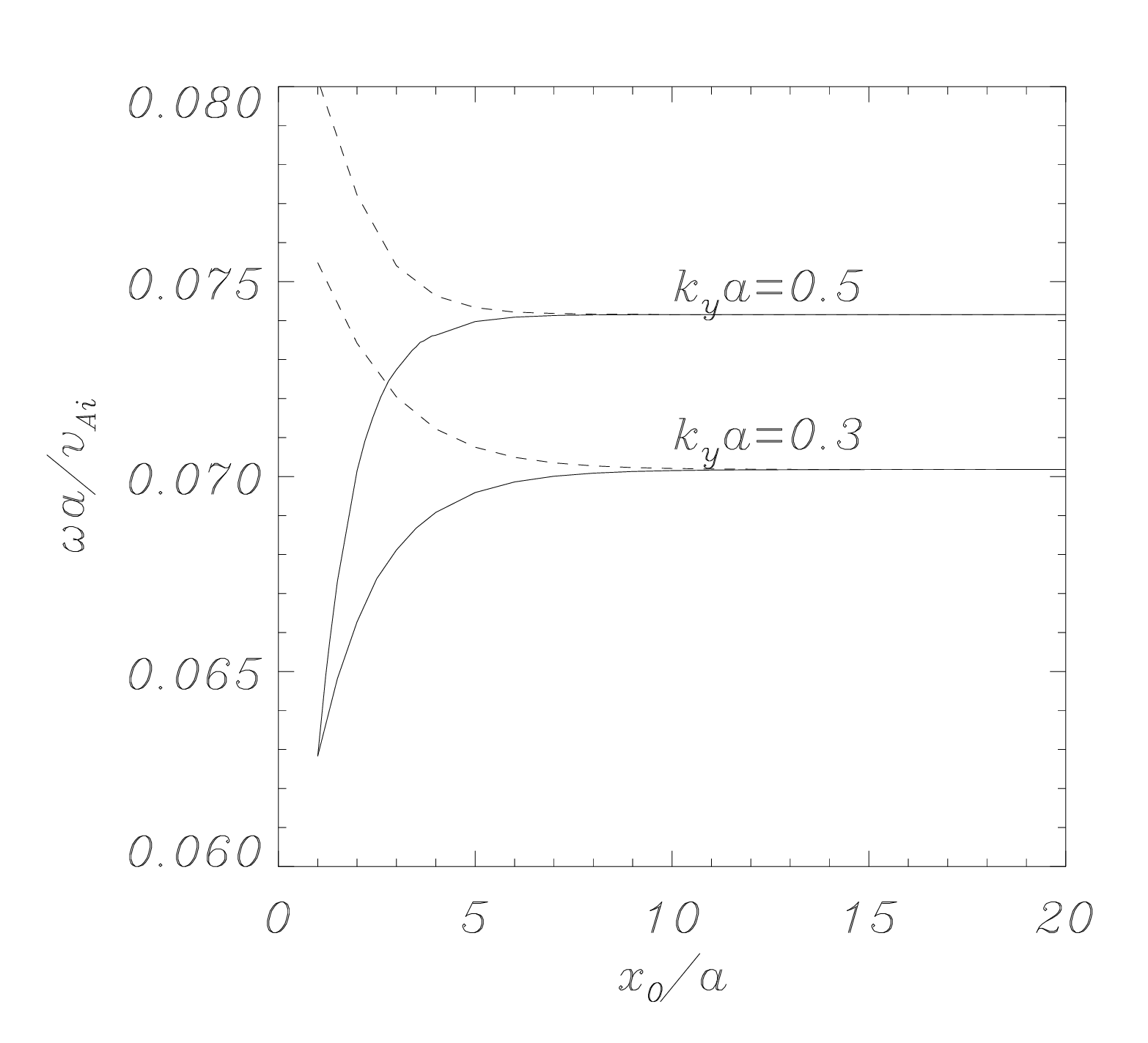}
   \includegraphics[width=7.0cm,angle=0]{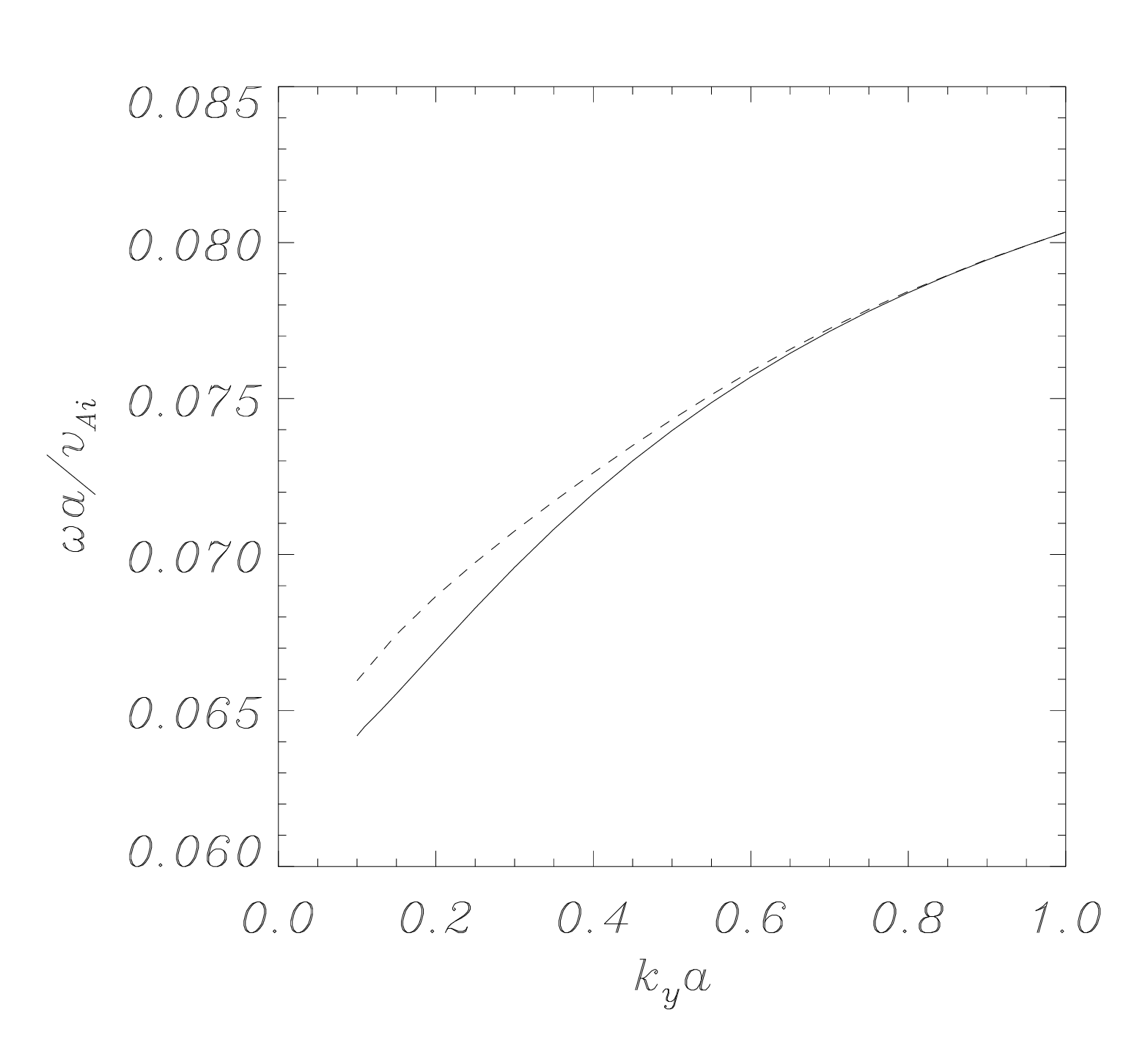}\\
   \includegraphics[width=7.0cm,angle=0]{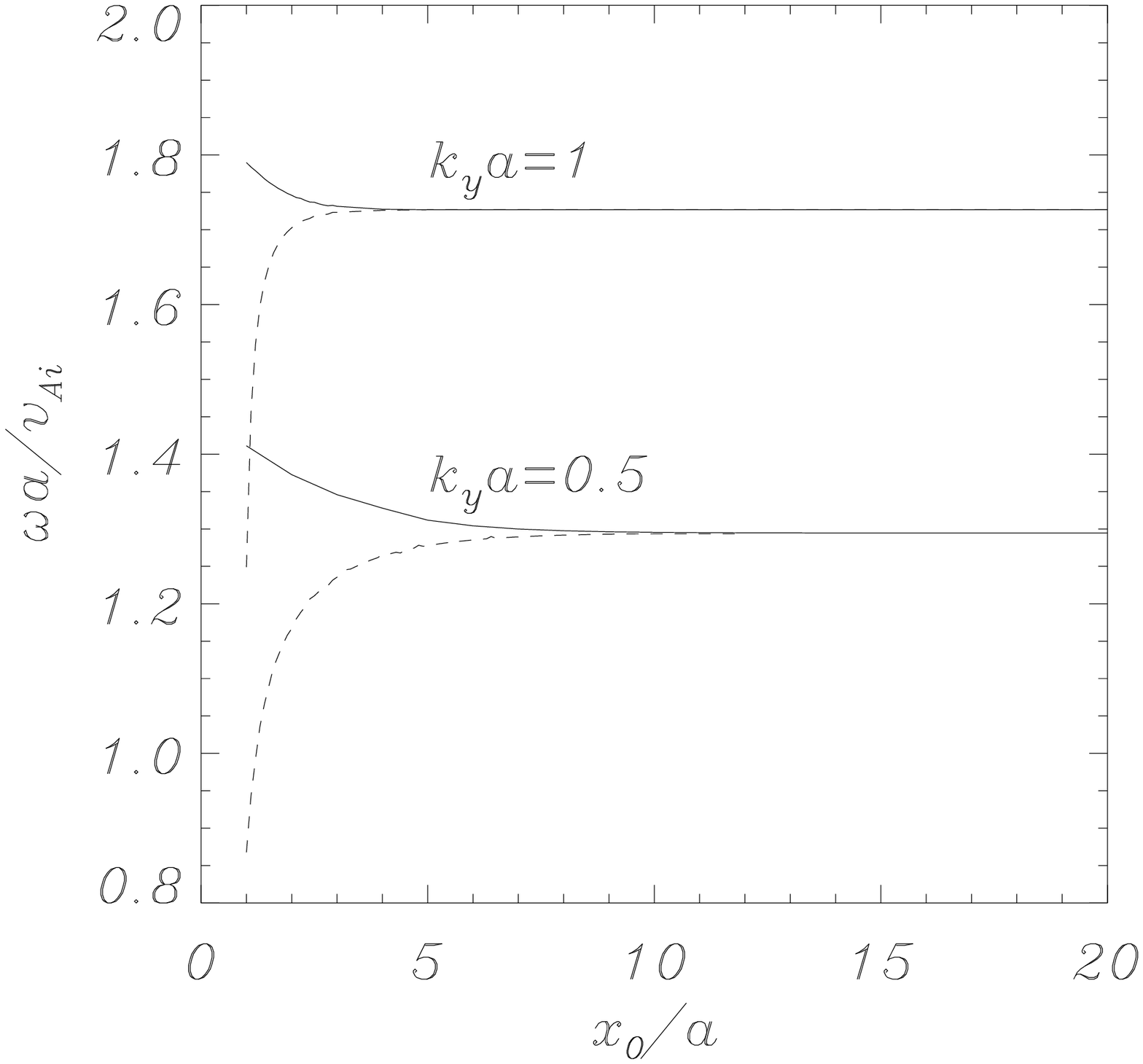}
   \includegraphics[width=7.0cm,angle=0]{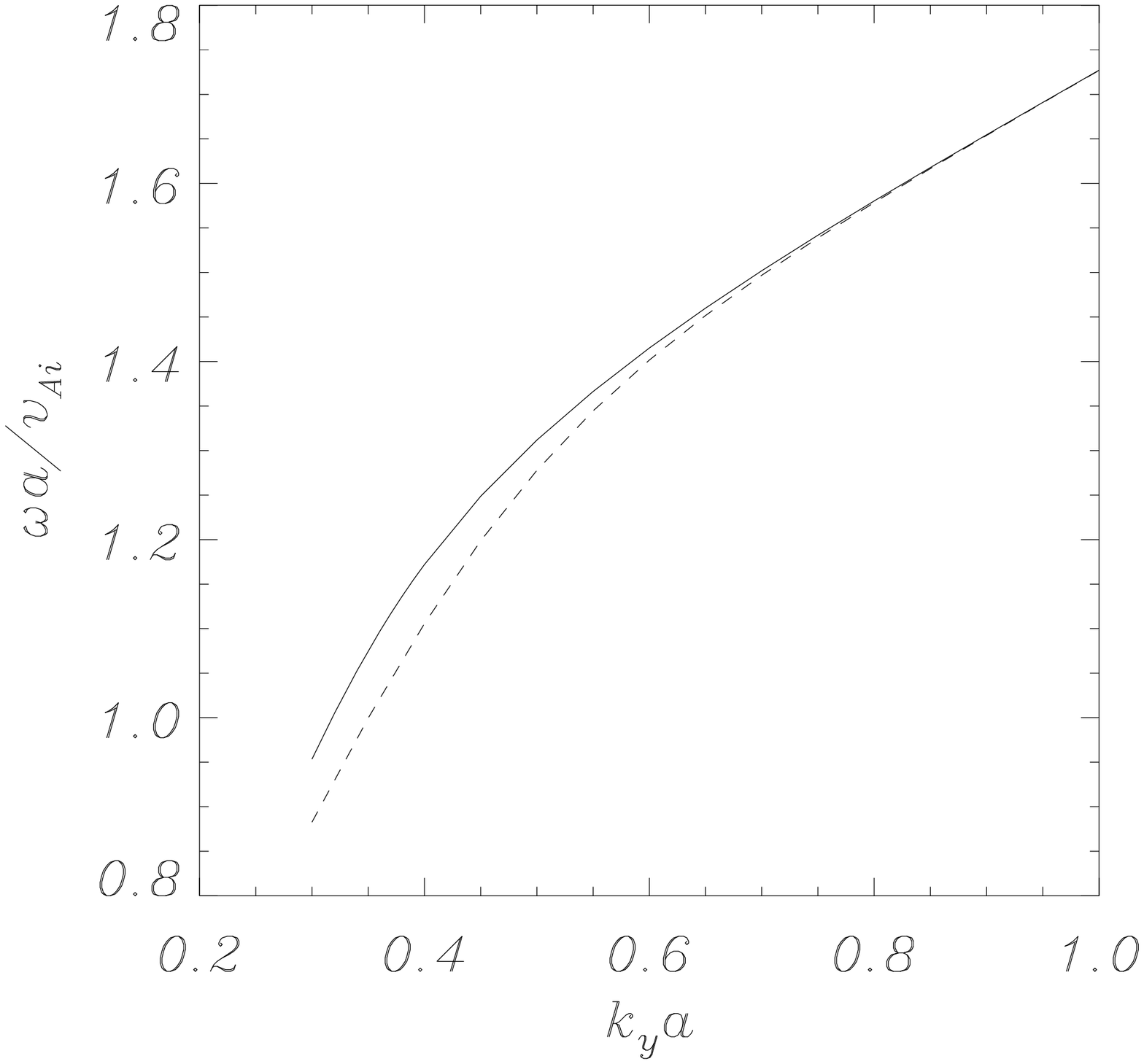}
     \caption{{\em Left}, Frequency as a function of the position of the slabs centers
for different values of the perpendicular
wavenumber for the kink solutions ({\em top}), 
the sausage surface solutions ({\em middle}), and the sausage body 
solutions ({\em bottom}). 
{\em Right}, Frequency as a function of the perpendicular wavenumber 
for the same six types of solution, for a fixed distance between the slabs, $x_0=5a$. In all the 
figures the solid lines correspond to the symmetric solution and the dashed lines to the antisymmetric
solution.}
\label{freq}
\end{figure}

When finding our solutions, we consider fixed
values for the density contrast of the two-slab
system, $\rho_i/\rho_e$=$10$, and for the longitudinal wavenumber,
$k_za$=$\pi/50$. For the observed kink oscillations
with a wavelength double the length of the loop,
this corresponds to a ratio of length to width
$L/2a=25$. These are typical values for observed coronal
loops for which transverse oscillations have been reported.
We have studied the dependency of the frequency of these solutions as a function 
of the position of the slabs, related to the distance between the loops, and as a function of the 
perpendicular wavenumber. 
Figure~\ref{freq} displays the obtained results. 
The splitting of the solutions can clearly be seen for the six modes. For the kink symmetric 
and antisymmetric solutions, the range of distances for which the interaction is strong and the 
deviation from the kink mode frequency of a single slab significant, is large in the absence of perpendicular propagation 
($k_y=0$). 
Even for distances of the order of $100a$, the two solutions do not converge to 
that frequency. When perpendicular propagation is included the single slab kink mode frequency 
is approached for much smaller distances between slabs, in such a way that only for distances of the order of 
a few times $a$ the symmetric and  antisymmetric solutions have frequencies that differ significantly 
from the kink frequency of a single slab. Also,
the  increase of the perpendicular wavenumber 
allows the existence of the kink antisymmetric mode for smaller distances between slabs, 
in comparison to the $k_y=0$ case, when it becomes leaky below $x_0\simeq 26a$. 
As for the sausage surface and sausage body, symmetric and antisymmetric solutions, a similar result 
is obtained. They are plotted only for values of $k_y\neq0$ because the surface solutions only exist 
when $k_y\neq0$ and the body solutions are leaky
below a certain value of $k_y$. We can also appreciate the splitting of the single slab solutions and 
the dependency of the interaction between the slabs and, hence, the deviation from the single slab frequencies
as a function of the slab separation and
perpendicular wavenumber. It is interesting to note that the fundamental 
solution in the case of kink and sausage surface solutions is the symmetric one, 
while it is the antisymmetric one in the case of the sausage body solutions. 
The right panels of Figure~\ref{freq} display the frequencies of the six solutions as a function of 
$k_y$ for a fixed value of the distance between the slabs.
As we increase the value of the perpendicular
wavenumber both symmetric and antisymmetric solutions converge 
to the same frequencies  and approach those of the kink, sausage surface, and sausage body solutions of a 
single slab in the limit of quasi-perpendicular propagation.

We have computed the difference between the frequency of the  symmetric and antisymmetric solutions of the two-slab
system and the kink and sausage solutions of a single slab. For the kink solutions, and considering for 
example $x_0=2a$, the frequency of the symmetric mode is 15\% smaller than the kink mode frequency of 
a single slab, when $k_y=0$. This difference decreases to 3\% for $k_y a=1$. When the two-slab system
is oscillating in the antisymmetric mode, its frequency is 13\% larger that the kink mode frequency for 
$k_y a=0.5$, while it is only 4\% for $k_ya=1$. As for the sausage surface solutions of the two-slab system, 
a deviation from the sausage surface frequency of a single slab around 5\% is found, in the case of the
symmetric solution with $k_ya=0.3$ and $x_0=2a$. Finally, for the sausage body 
solutions of the two-slab system with $x_0=2a$, the maximum difference occurs for the antisymmetric solution with 
$k_ya=0.5$ and is around  10\%.

\begin{figure}
 \includegraphics[width=8.0cm,angle=0]{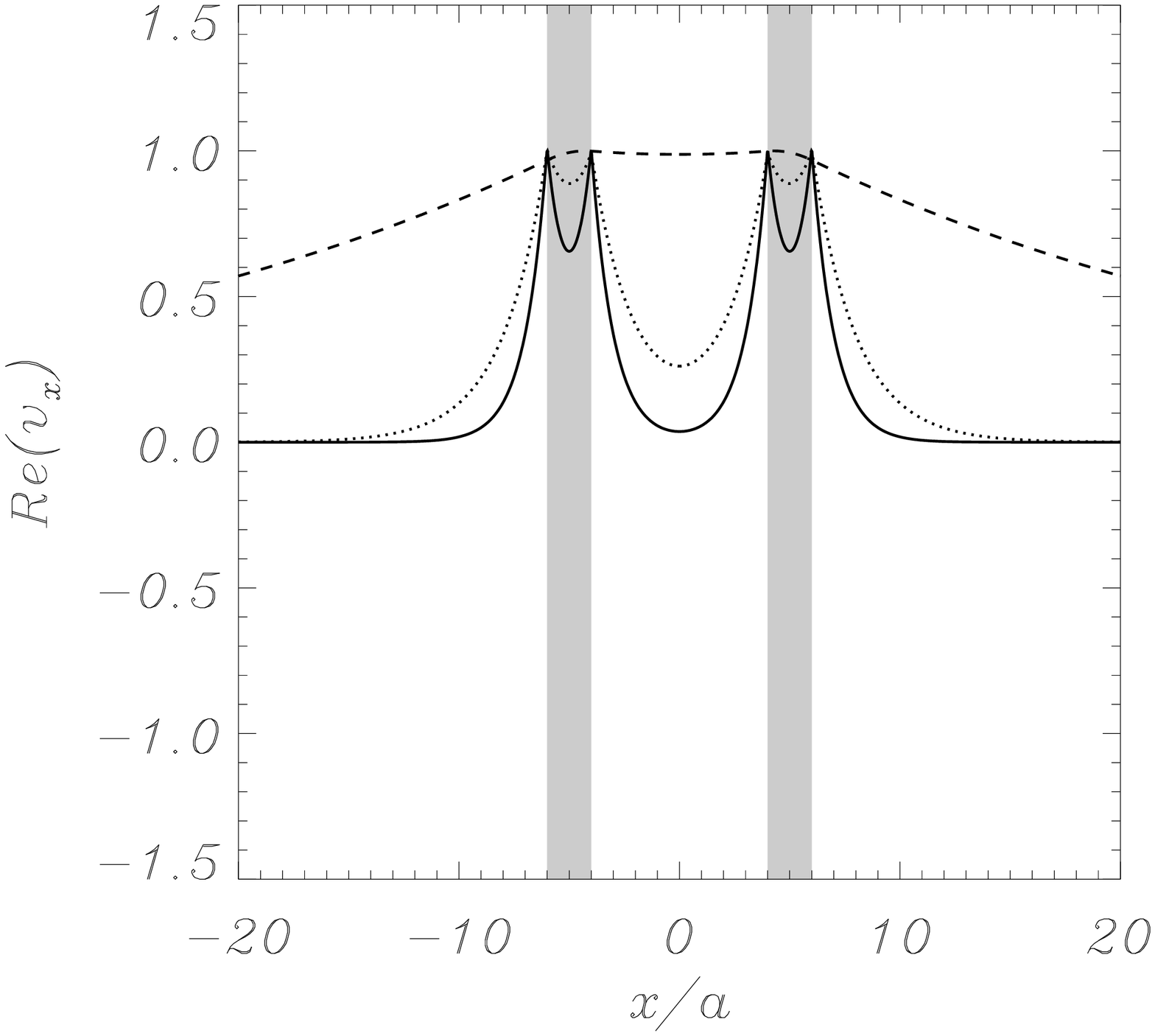}
\includegraphics[width=8.0cm,angle=0]{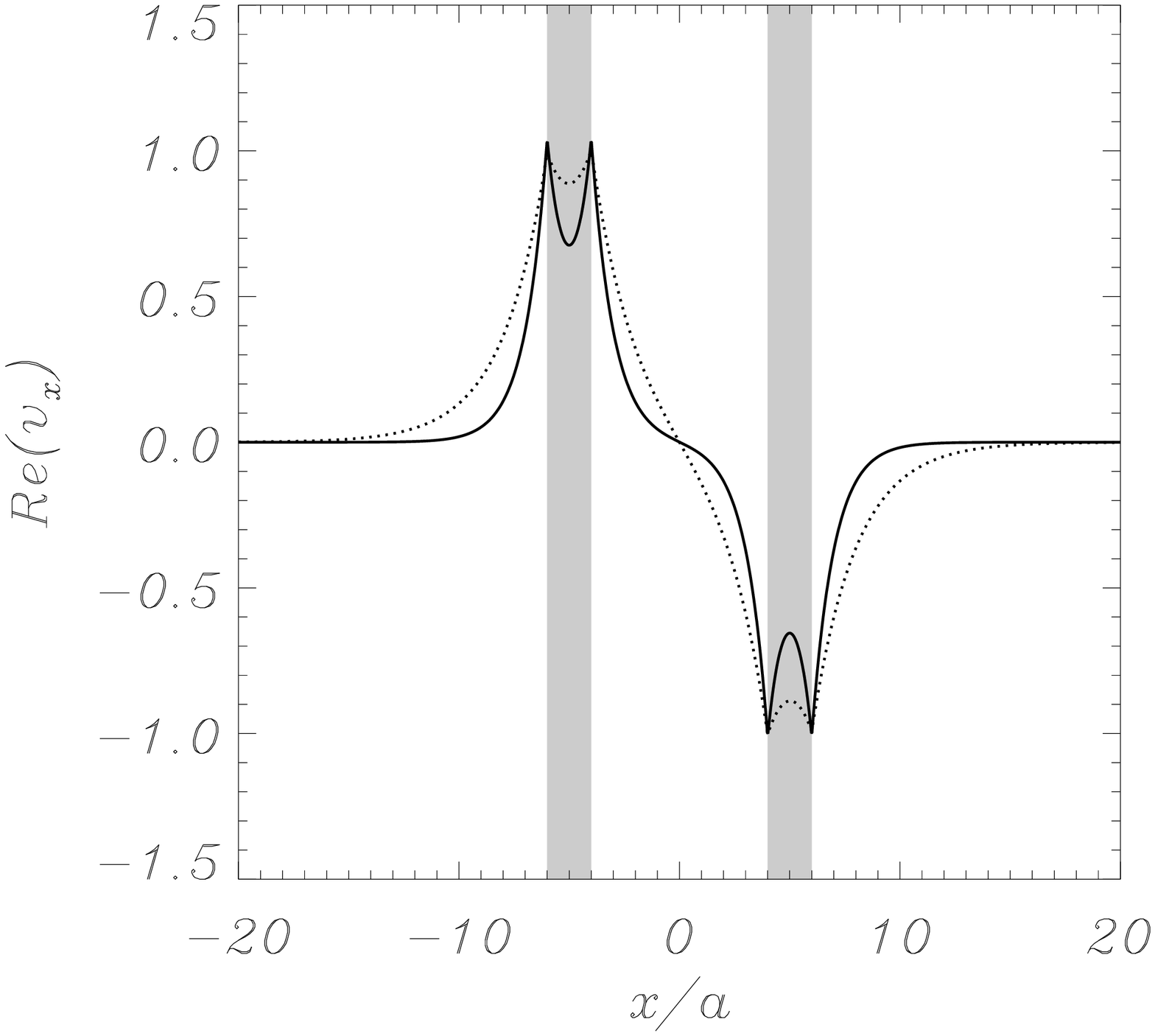}\\
\includegraphics[width=8.0cm,angle=0]{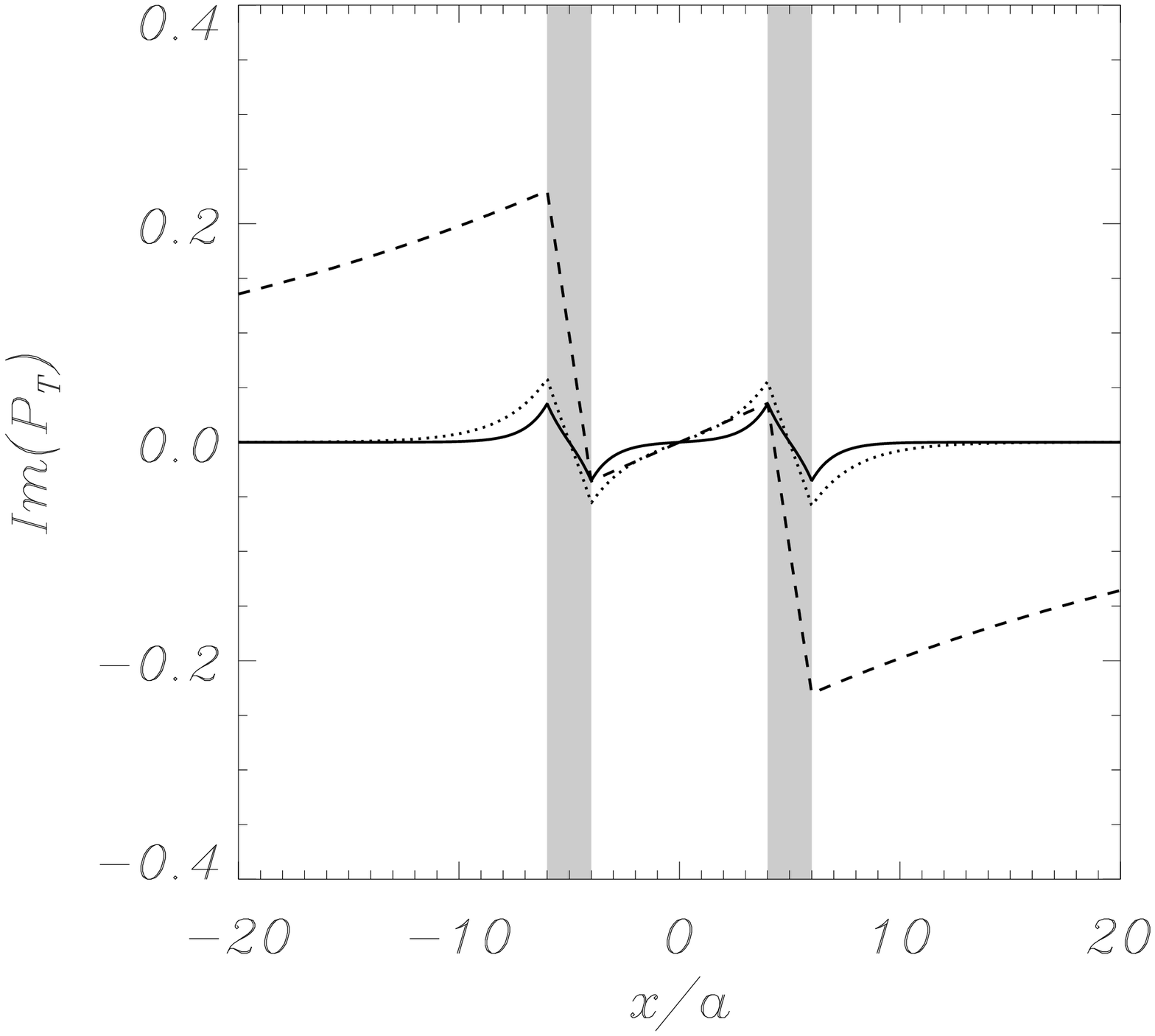}
\includegraphics[width=8.0cm,angle=0]{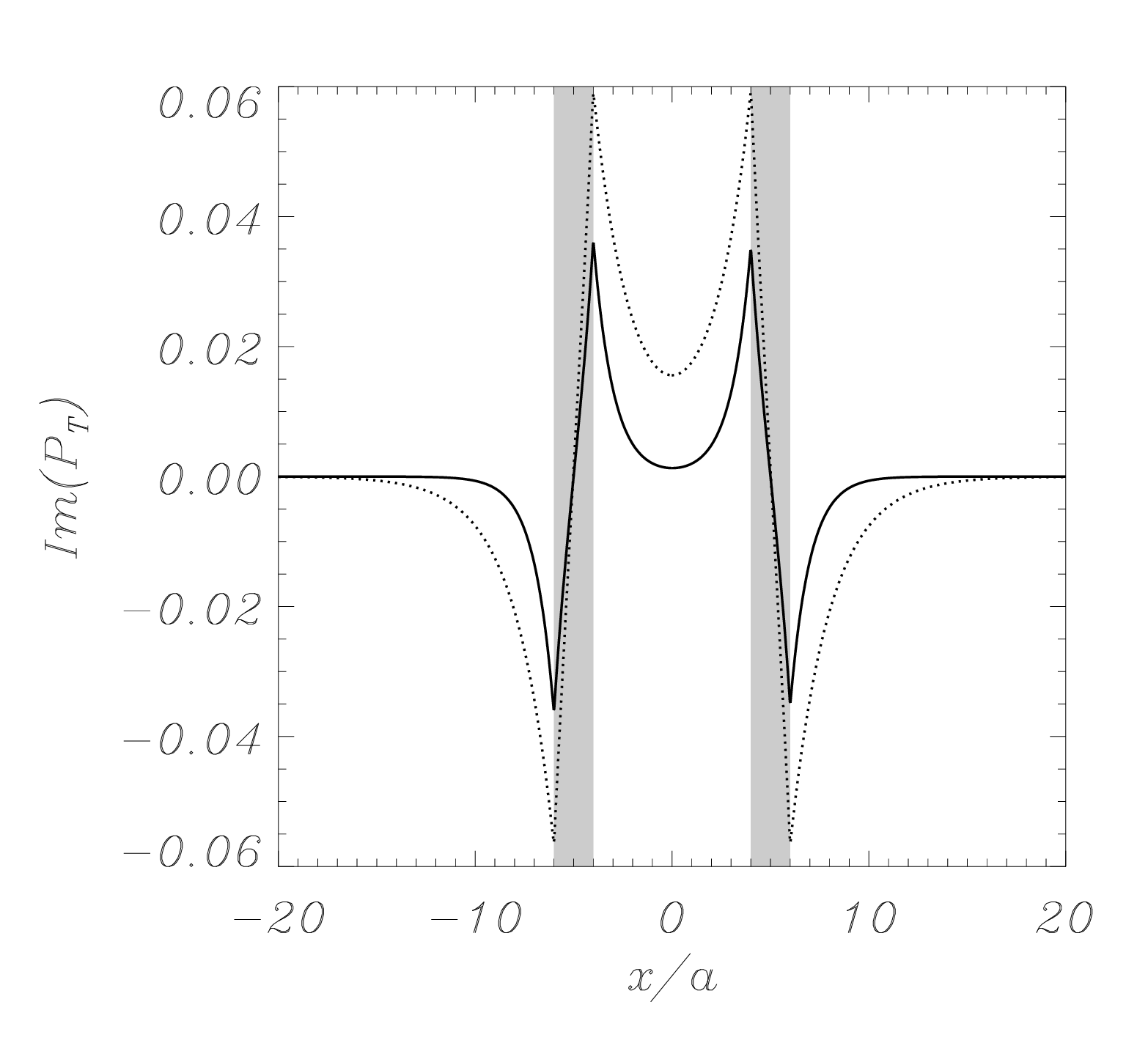}\\
\caption{Real part of the transversal velocity component, $v_x$, and imaginary part of the 
perturbed total pressure, $P_T$, for the fundamental symmetric ({\em left}) and antisymmetric 
({\em right}) kink solutions, for several values of the perpendicular wave number: $k_ya=0$ 
(dashed lines), $k_ya=0.5$ (dotted lines), and $k_ya=1$ (solid lines). Note that the 
character of the eigenfunctions for the symmetric mode changes from body-like to 
surface-like solution as $k_y$ is increased. In all figures the light-shaded regions represent the density enhancements.
These solutions have been obtained with a uniform computational grid with $N_x=10\ 000$ points in the 
range $-80\leq x/a \leq 80$ (except for the kink solutions with $k_y=0$ for which a range $-200\leq x/a 
\leq 200$ has been used).}
\label{kinkkink}
\end{figure}

\begin{figure}
 \includegraphics[width=8.0cm,angle=0]{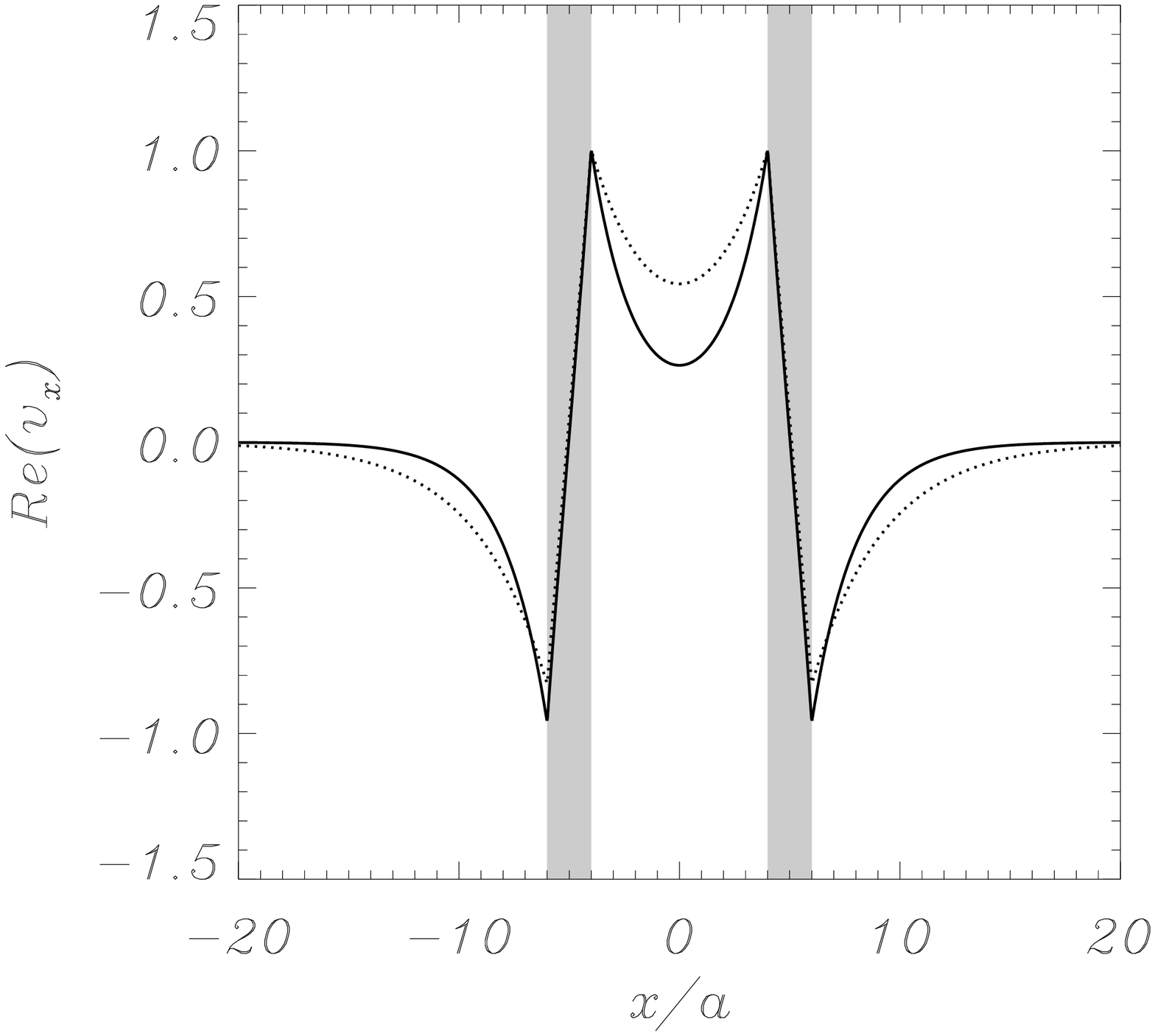}
\includegraphics[width=8.0cm,angle=0]{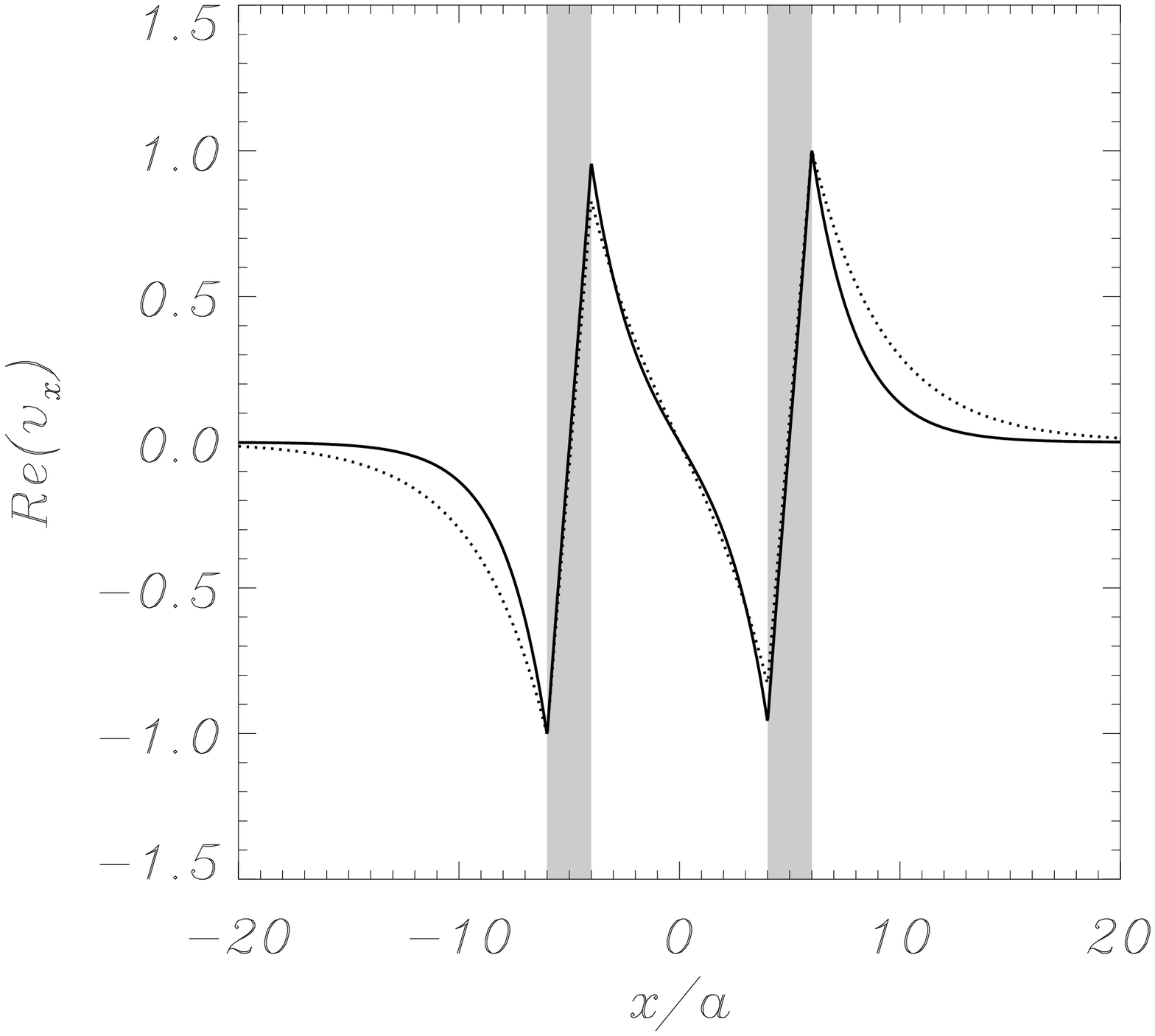}\\
\includegraphics[width=8.0cm,angle=0]{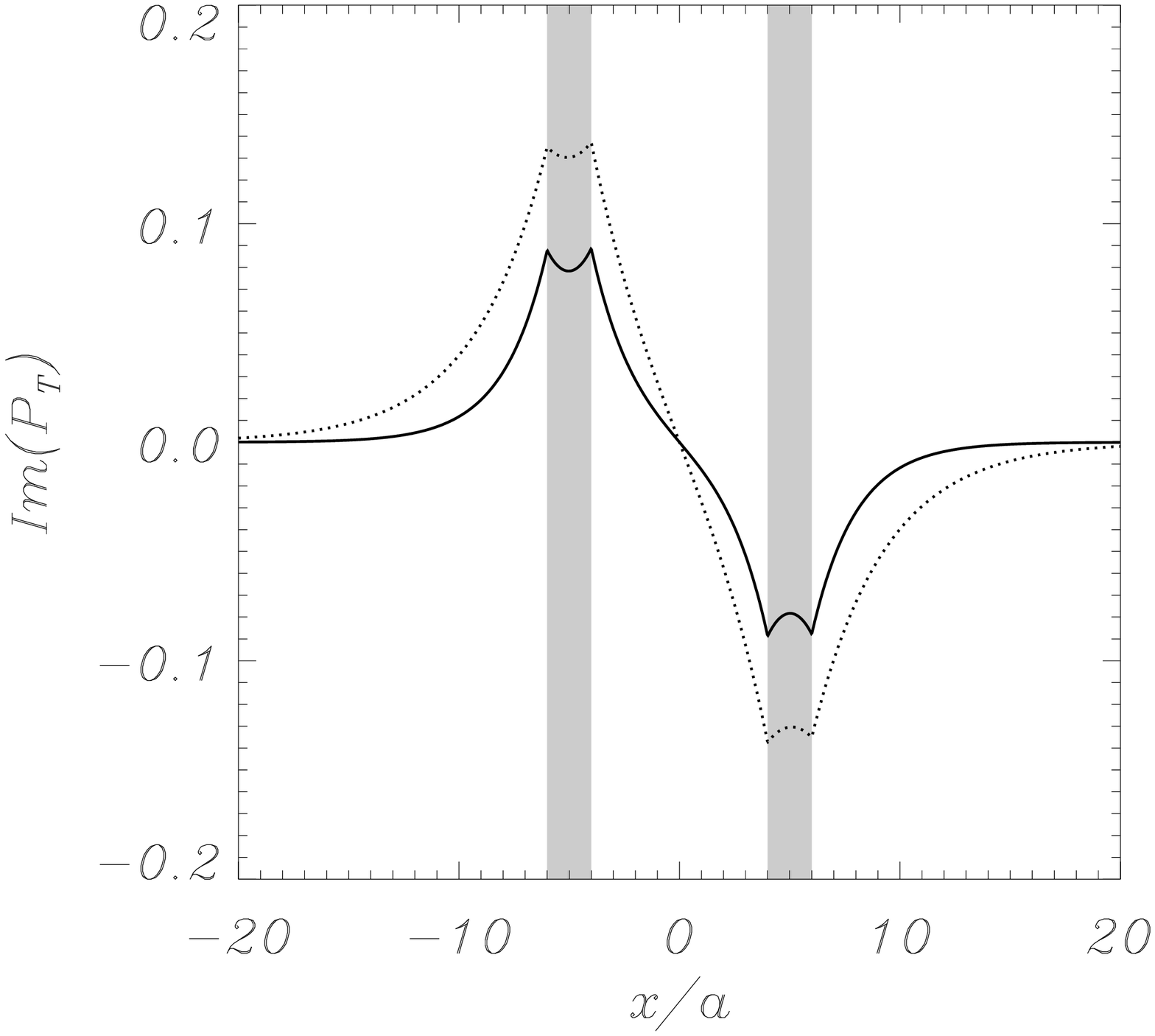}
\includegraphics[width=8.0cm,angle=0]{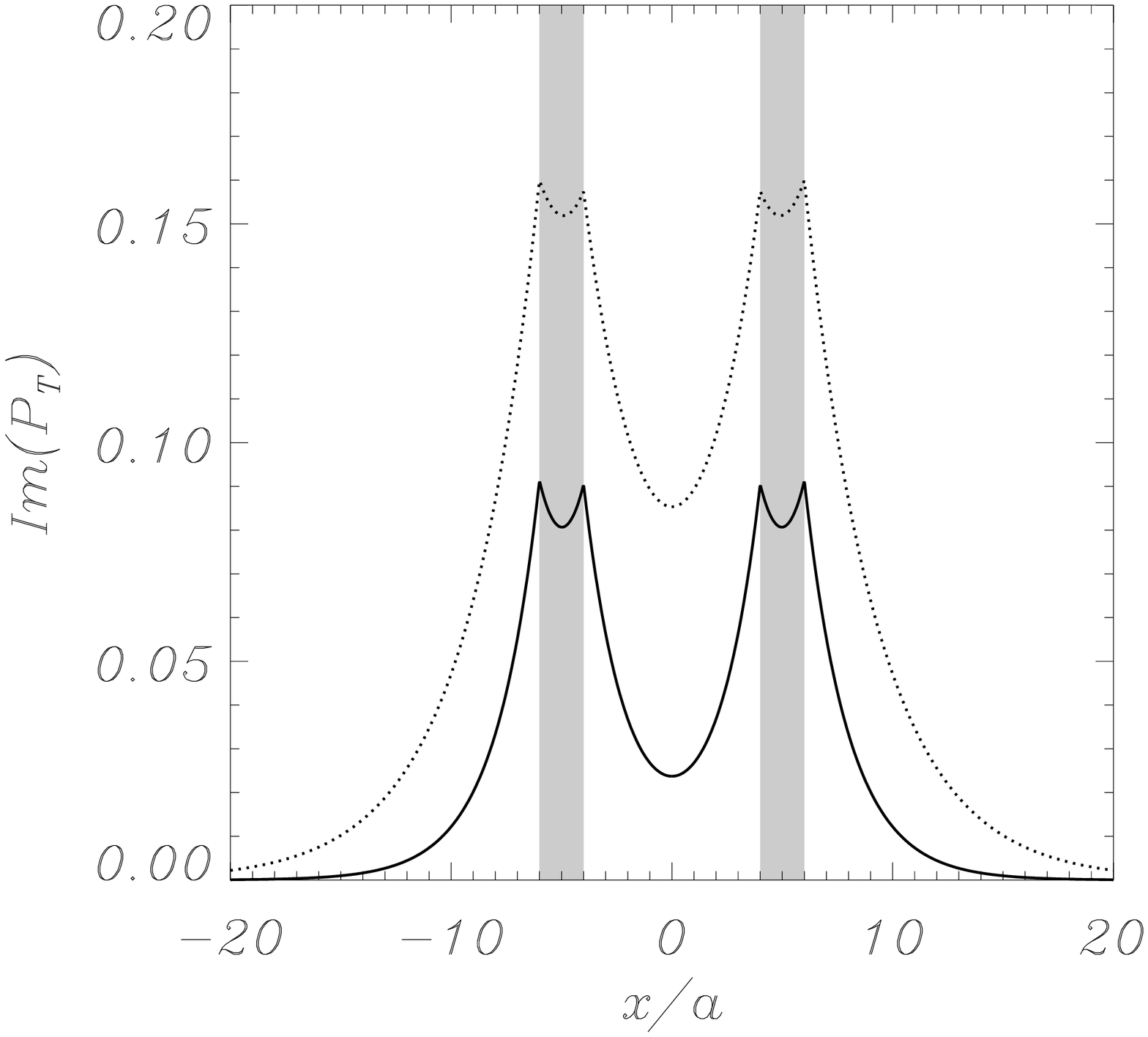}
\caption{Same as Figure~\ref{kinkkink} for the sausage surface, symmetric ({\em left}) and 
antisymmetric ({\em right}) solutions and two
values of the perpendicular wavenumber: $k_y a=0.3$ 
(dotted lines) and $k_y a=0.5$ (solid lines). These solutions are surface-like for all values of the 
perpendicular wavenumber.} 
\label{ssss}
\end{figure}

\begin{figure}
 \includegraphics[width=8.0cm,angle=0]{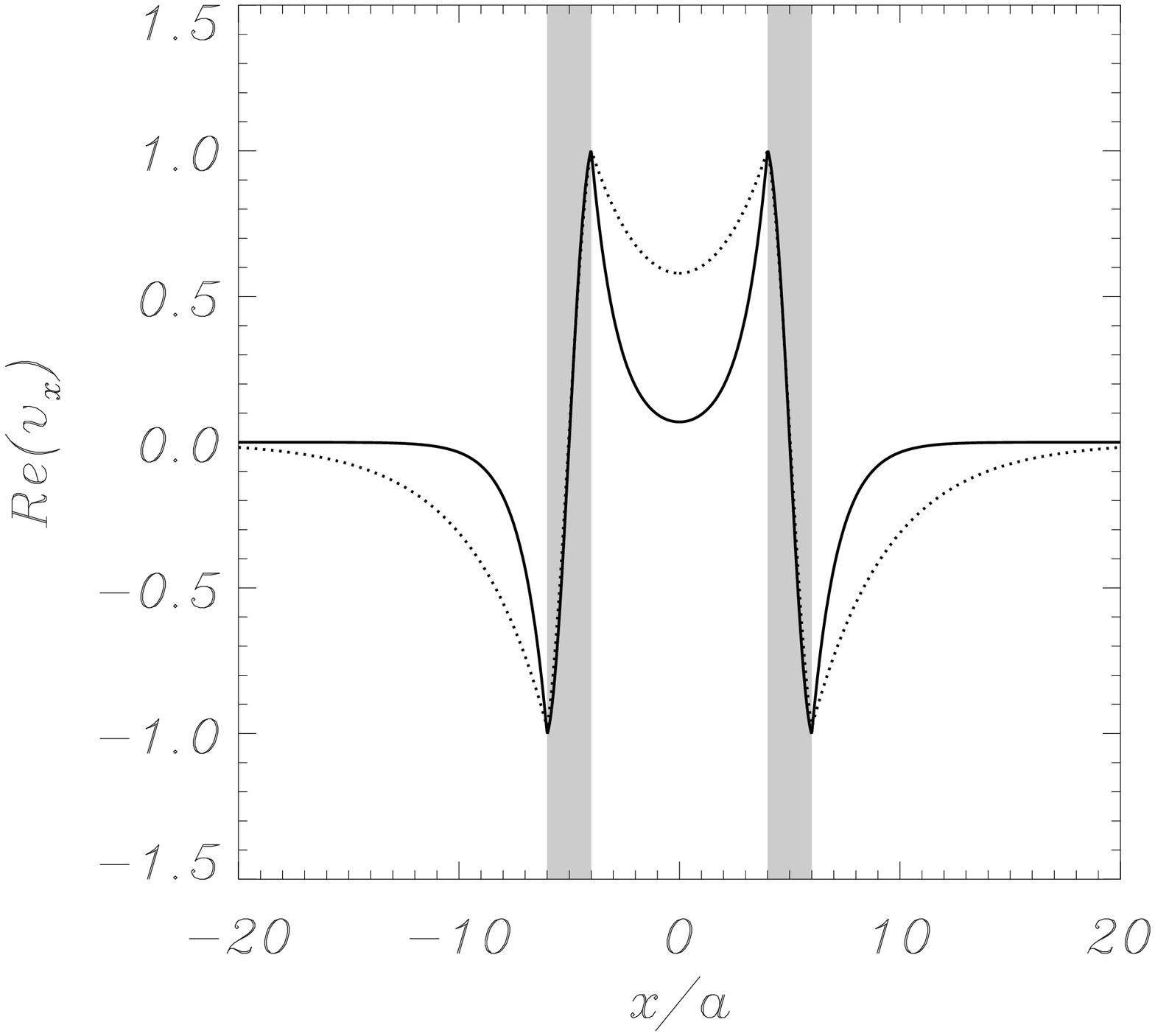}
\includegraphics[width=8.0cm,angle=0]{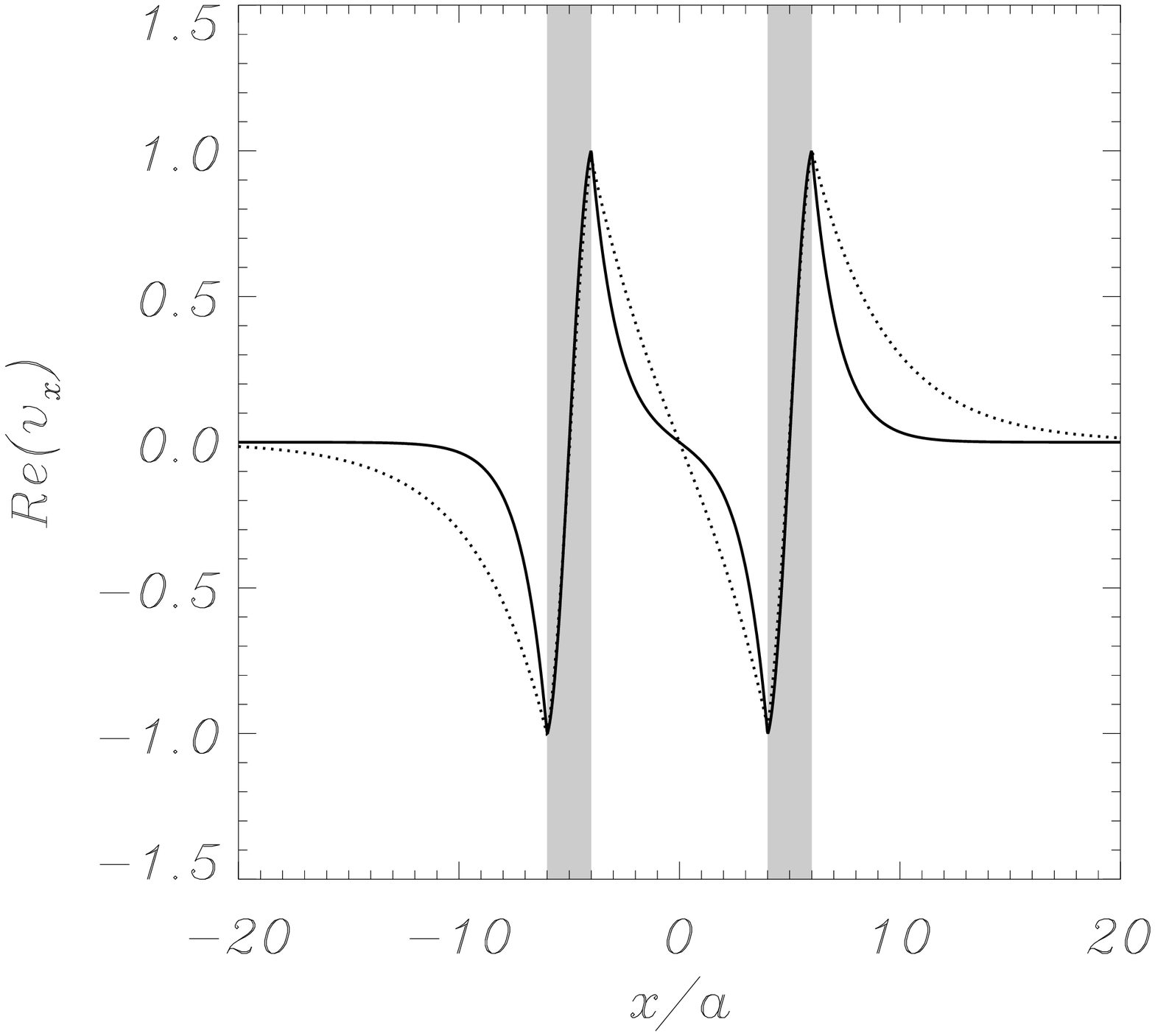}\\
\includegraphics[width=8.0cm,angle=0]{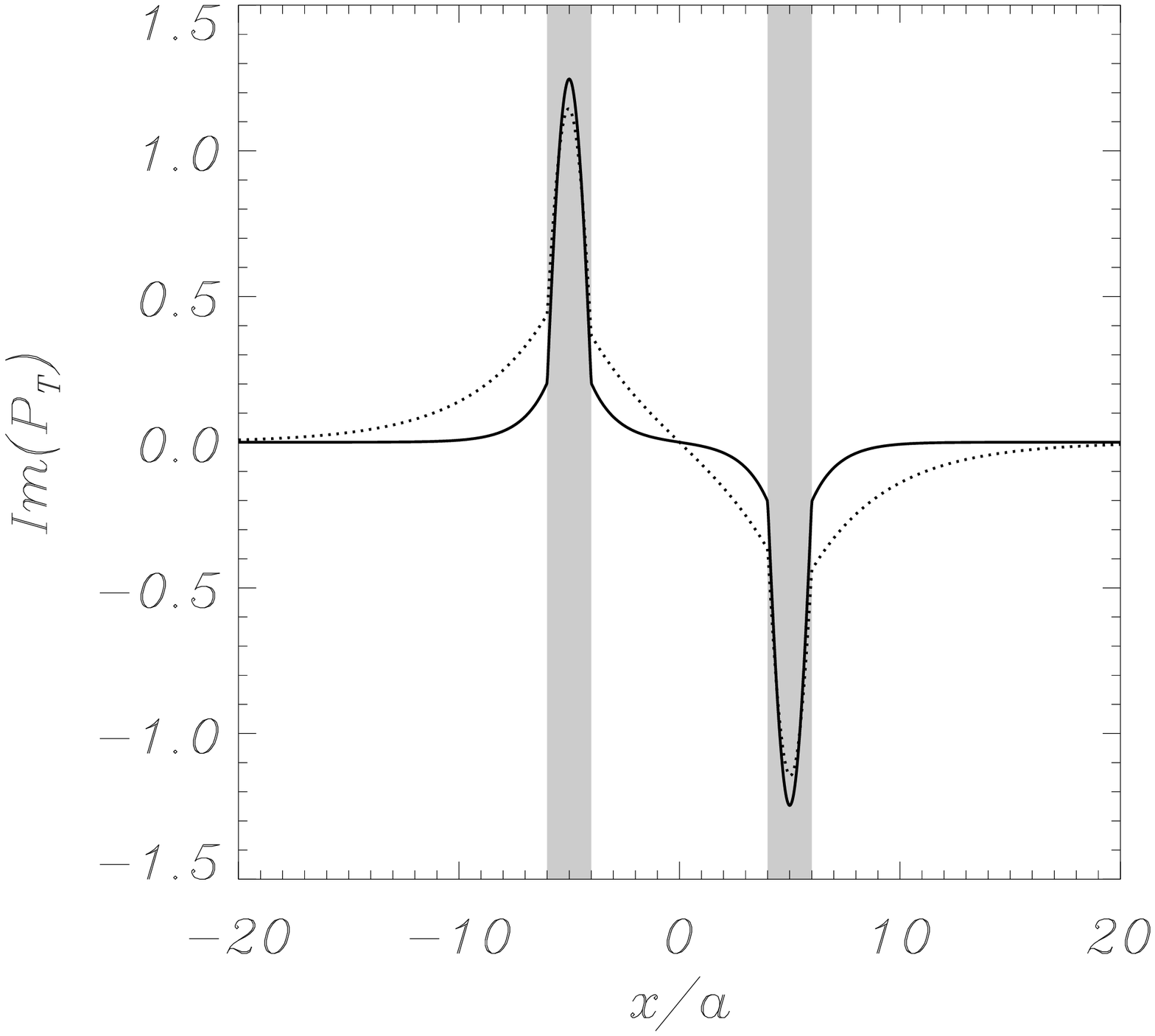}
\includegraphics[width=8.0cm,angle=0]{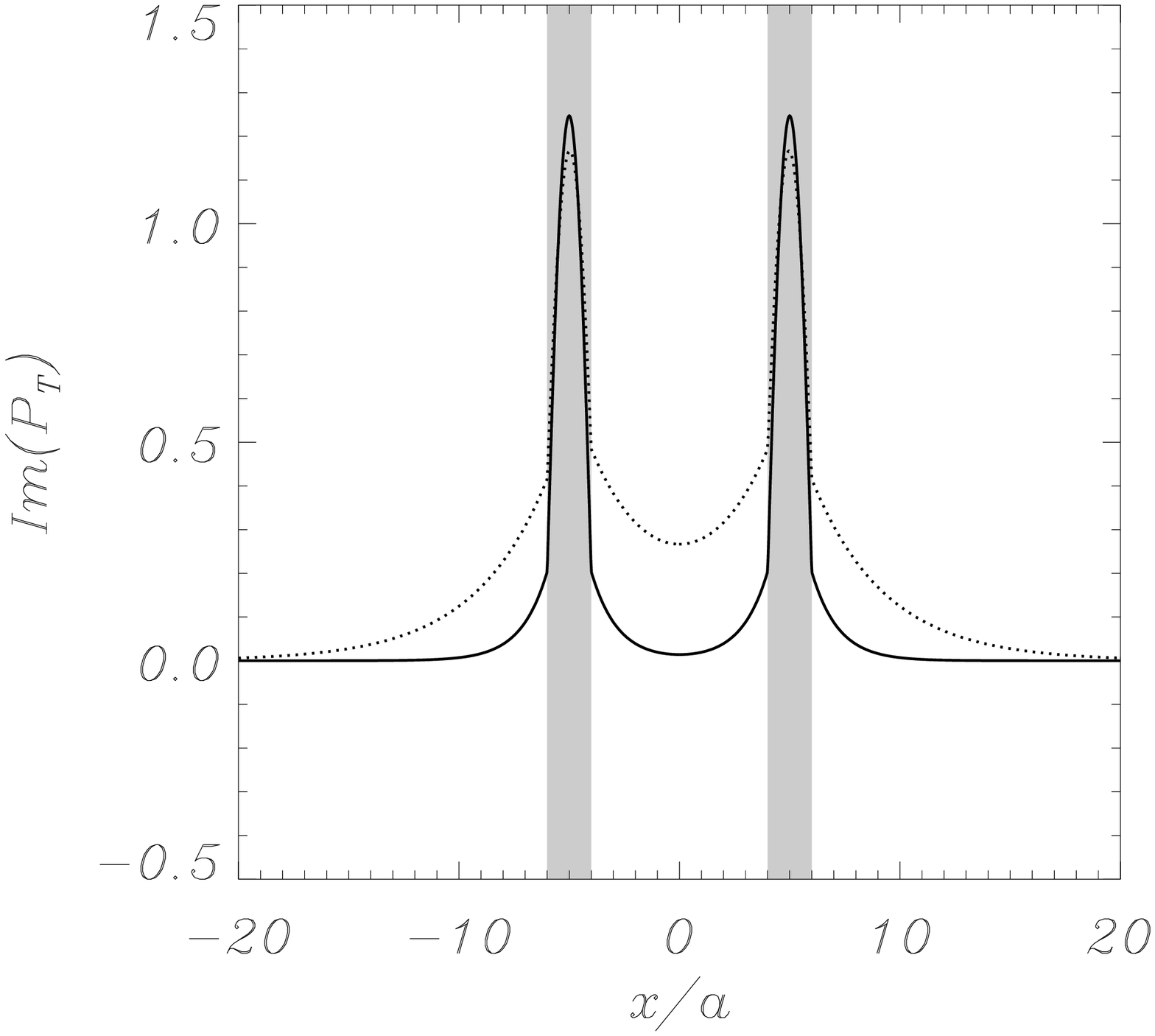}\\
\caption{Same as Figures~\ref{kinkkink} and \ref{ssss} for the sausage body, symmetric ({\em left}) and 
antisymmetric ({\em right}) solutions and two
values of the perpendicular wavenumber: $k_y a=0.5$ (dotted lines) 
and $k_y a=1$ (solid lines).
Note that for these solutions $v_x$ is very similar to the ones depicted in Figure~\ref{ssss}, 
while the pressure perturbation in and around the slabs differ substantially.
These solutions are body-like for all values of the perpendicular wavenumber.}
\label{sbsb}
\end{figure}

The results displayed in Figure~\ref{freq} can easily be interpreted by analyzing the spatial distribution
of the corresponding eigenfunctions. Figures~\ref{kinkkink}, \ref{ssss}, and \ref{sbsb}  show the 
transversal component of the perturbed velocity and the total pressure perturbation for the kink, sausage 
surface, and sausage body, symmetric and antisymmetric solutions, for different values of the 
perpendicular wavenumber. In Figure~\ref{kinkkink} the solutions for the kink solutions are plotted for three 
values of the perpendicular wave number. For $k_y=0$, the transversal velocity is maximum at the internal part 
of the slabs, while it decreases exponentially outside. The total pressure perturbation has an antisymmetric profile in 
each slab, with a zero value at the center. When oblique propagation is included, there is an improved confinement of 
the eigenfunctions with a sharper drop-off rate in the external medium, a behavior also found by \citet{toni03}, in the context of 
prominence fibril oscillations. Also, the character of the solutions inside 
the slabs change, and the two solutions become surface-like solutions, with a decreasing amplitude of $v_x$ 
inside the slabs as $k_y$ is increased. This improved confinement of the solutions weakens the interaction between 
the slabs and results in a frequency of the two-slab kink  solution closer to that of the kink mode of a single slab. 
Regarding the sausage 
surface and body-like solutions (Figures~\ref{ssss} and \ref{sbsb}), we can appreciate that they have a 
similar spatial distribution of eigenfunctions. The main difference has to be found in the total pressure perturbation 
profiles, that are maximum at the center of each slab, for the sausage body solutions, while they peak at 
the edges of the slabs for the surface-like counterparts.
An increase in the perpendicular wavenumber also produces an improved confinement, in the external regions, 
for these four types of modes. This also helps explaining why the interaction between the slabs weakens for 
increasing perpendicular wavenumber. However, in contrast to the kink solutions, now the transversal velocity
component inside the slabs is not affected by the increase in $k_y$ and the solutions keep their surface or body character. 
Finally, oblique propagation of waves affects 
the magnitude of the total pressure perturbation of the sausage surface, symmetric and antisymmetric solutions 
(Figure~\ref{ssss}), but not very much that of the sausage body, symmetric and antisymmetric, modes inside the 
density enhancements (Figure~\ref{sbsb}).

In summary, the inclusion of perpendicular propagation of perturbations changes significantly the oscillatory properties
of fast MHD waves in our two-slab system. First, it allows the existence of two sausage surface solutions that are not 
present when $k_y=0$. Also, it allows the existence of trapped sausage body solutions for slab separations 
for which they are leaky when $k_y=0$. Second, it produces an improved confinement of the solutions that, for the 
six modes, are better confined to the slabs, with a sharper drop-off rate in the coronal medium. This produces 
changes in the frequencies of the six modes and varies significantly the range of slab separations for which the 
difference between the symmetric and antisymmetric frequencies and the single slab frequencies are important. In the 
case of the kink solutions, the increase of the
perpendicular wavenumber also changes the character of the solutions 
inside the slabs, and the modes  become
surface-like. Therefore, for a fixed value of the
parallel wavenumber, both the 
separation between the loops and the value of the
perpendicular wavenumber determine the strength of the 
interaction between the slabs and the departure of the obtained frequencies from the results obtained with a single 
loop model.

\subsection{Resonant Damping of Oscillations}\label{damping} 


Observations of transversal coronal loop oscillations show that motions are rapidly damped in only 
a few periods \citep{Aschwanden99,Nakariakov99}. Coronal loops forming an arcade
also show this property, such as reported by \citet{Verwichte04}.
Resonant absorption of wave energy has been  proposed as a mechanism for the damping of loop oscillations
by \citet{HY88}, prior to observational evidence, provided by TRACE observations,
of damped transversal loop oscillations.
 \citet{GAA02} provided an interpretation to the damping observed in TRACE loop oscillations
based on this mechanism. Although there is still no consensus on what causes the rapid damping of loop 
oscillations, several authors have presented comparisons between theory and observations that point out 
that resonant damping is a viable mechanism \cite[see][for
example]{GAA02,Aschwanden03b,Arregui07}. 
This damping mechanism relies on the coupling of global fast motions and localized Alfv\'enic motions due to 
the non-uniformity of the medium in the transversal direction.


To study the damping of the normal modes described in \S~\ref{undamped}, we now include 
non-uniform transitional layers that connect the internal and external densities at the edges of each 
slab ($l\neq0$). For the normal mode analysis of Equations~(\ref{first})--(\ref{last}), a small but finite value 
of the resistivity has to be provided when computing the damping of oscillations.
The magnetic Reynolds number in the solar corona is believed to be of the order of $10^{14}$, hence the resistive diffusion time
is very large. Such large values of $R_m$ are difficult to handle in any numerical scheme. On the other hand,
as shown by \citet{POKE91}, the damping by resonant absorption becomes independent of resistivity, beyond a given value
of $R_m$. In our numerical computations,  resistivity has been chosen to be 
small enough for the imaginary part of the frequency to be independent of resistivity, so that we can be confident
that solutions that correspond to the solar coronal large Reynolds numbers are obtained. 
This condition has been checked to a high accuracy and for the solutions presented in this paper 
a value for the magnetic Reynolds number  $R_m=v_{Ai}a/\eta$ (with $v_{Ai}$ the internal Alfv\'en speed) 
in between $10^6$ and $10^8$ has been sufficient.

\begin{figure}
   \includegraphics[width=6.0cm,angle=0]{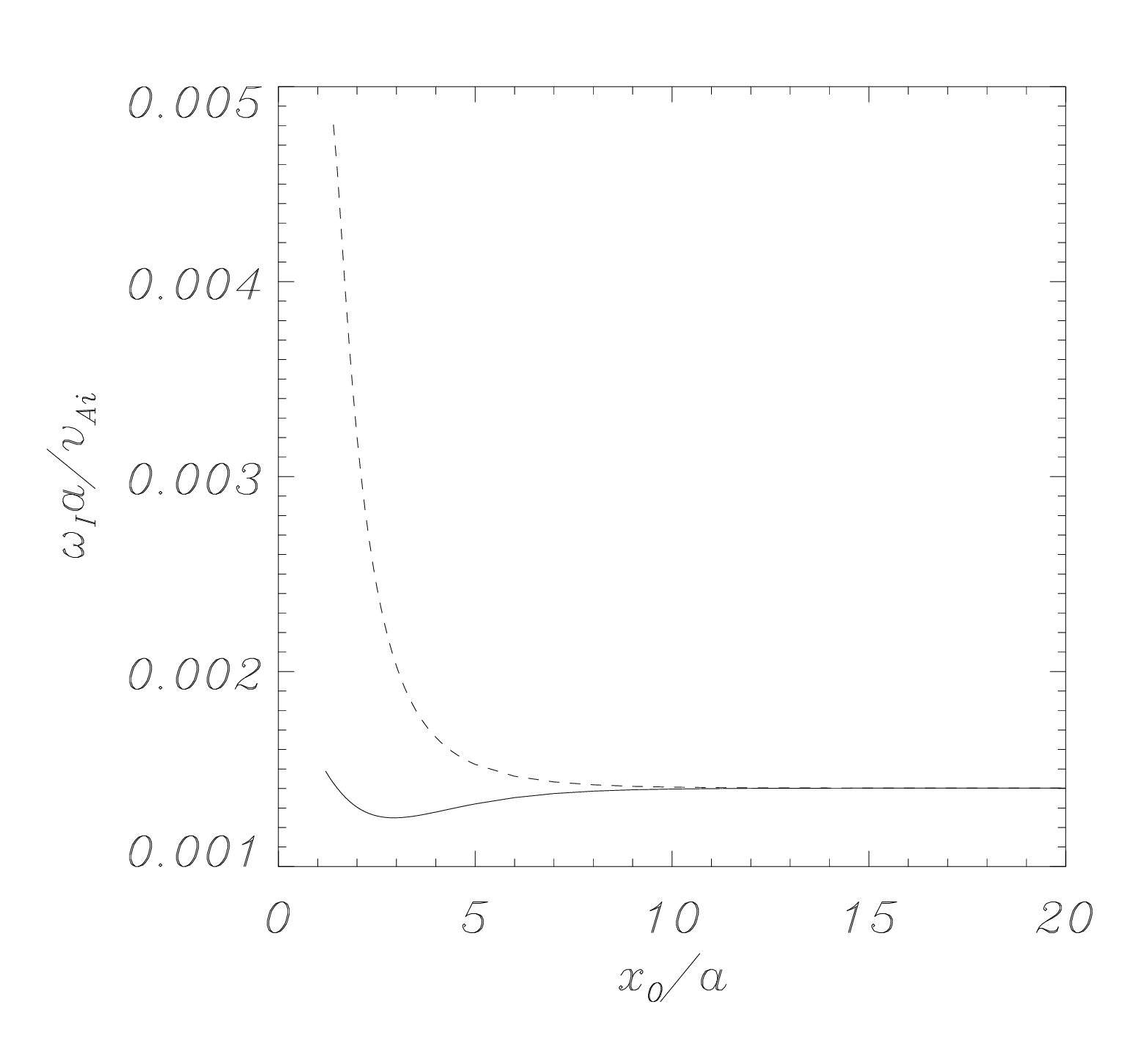}
   \includegraphics[width=6.0cm,angle=0]{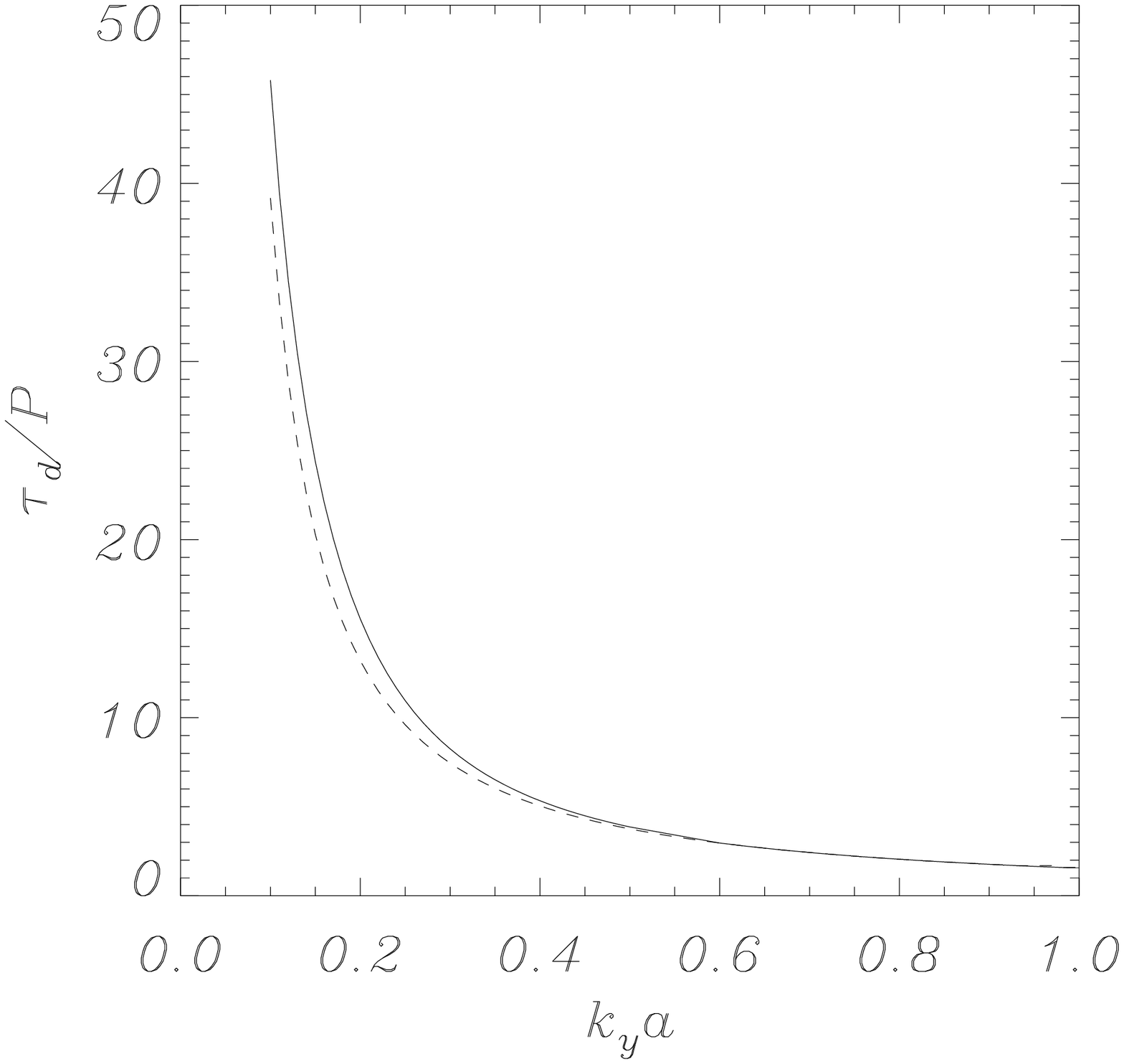}\\
    \includegraphics[width=6.0cm,angle=0]{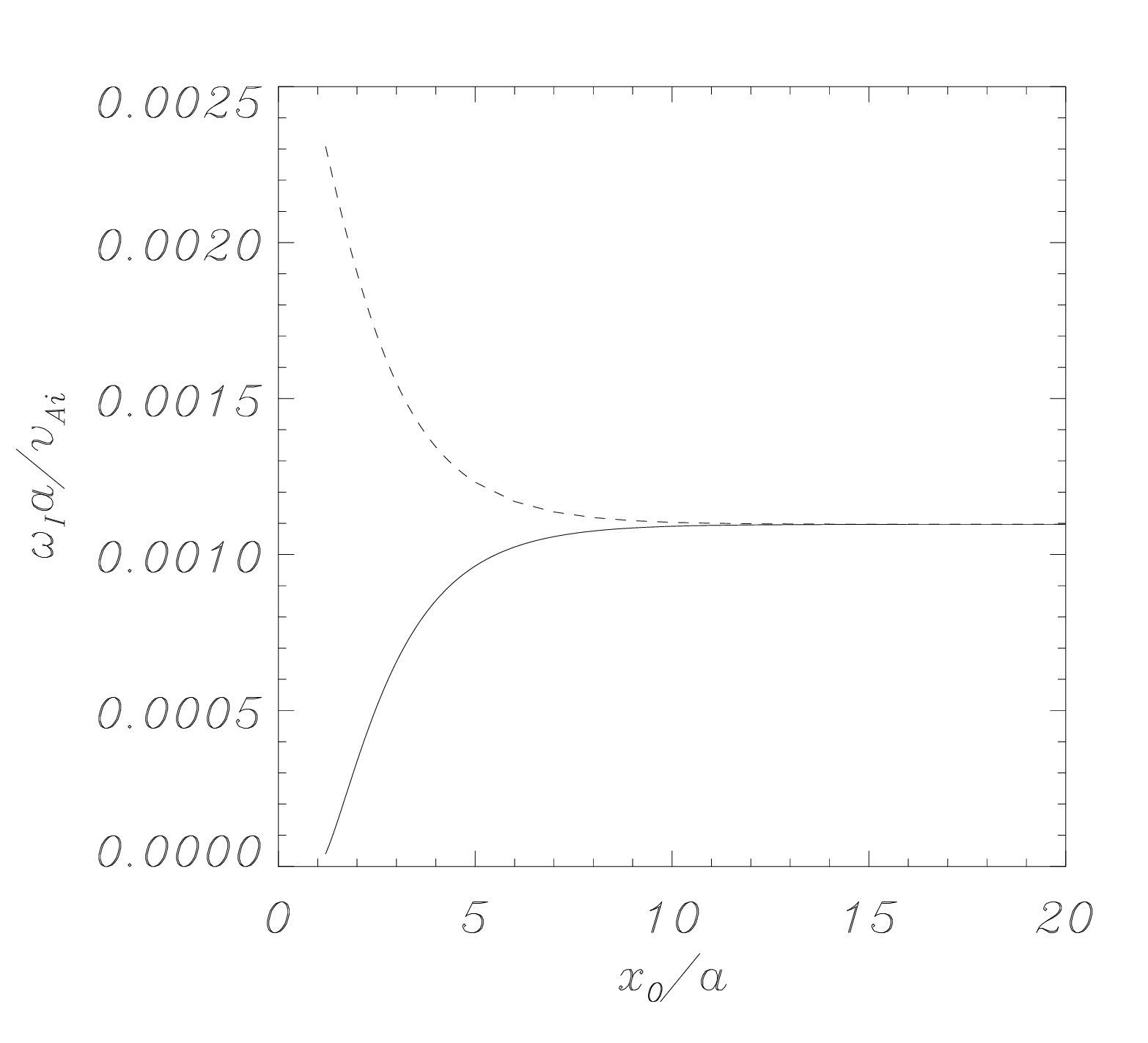}
    \includegraphics[width=6.0cm,angle=0]{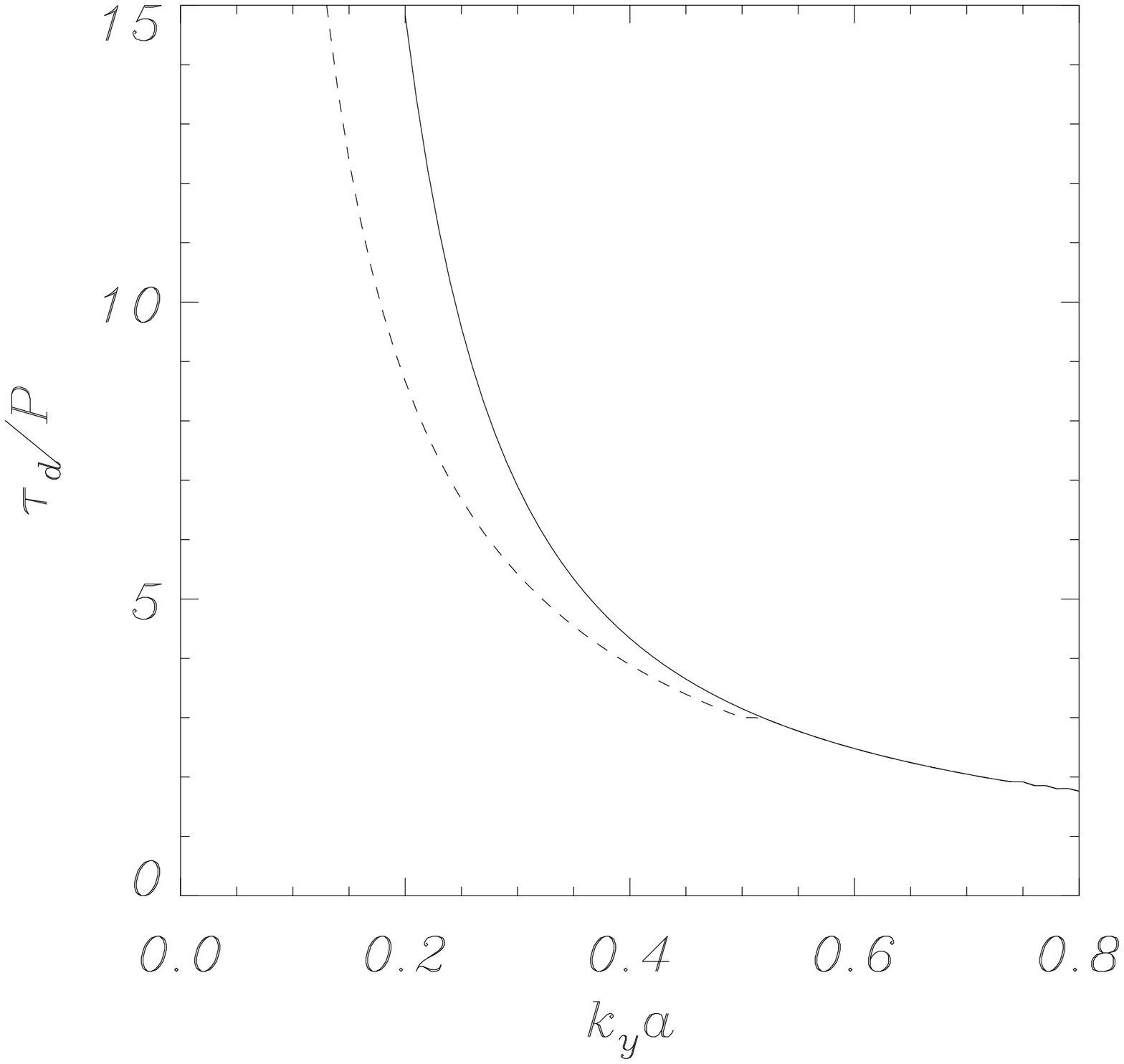}\\
    \includegraphics[width=6.0cm,angle=0]{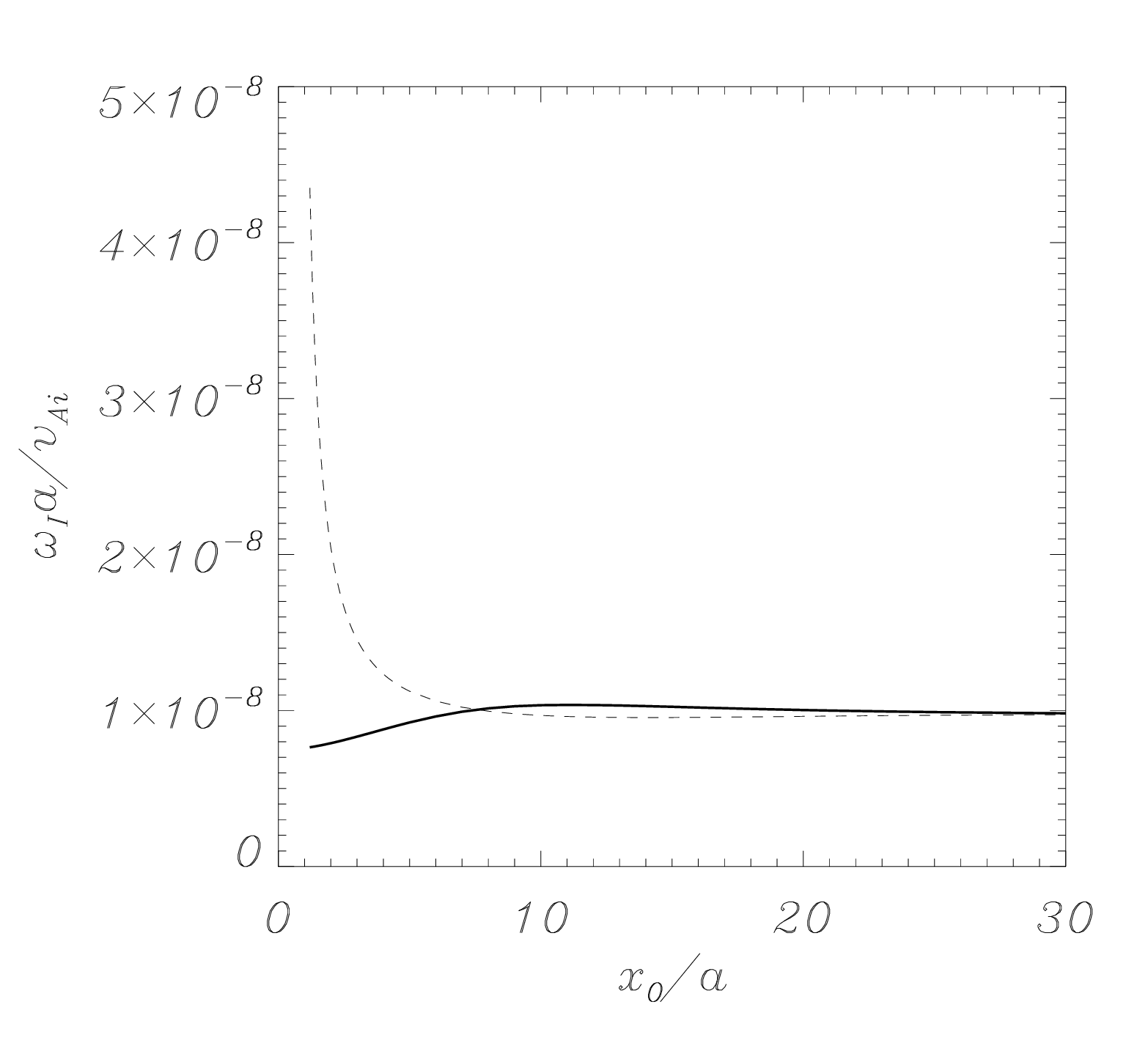}
    \includegraphics[width=6.0cm,angle=0]{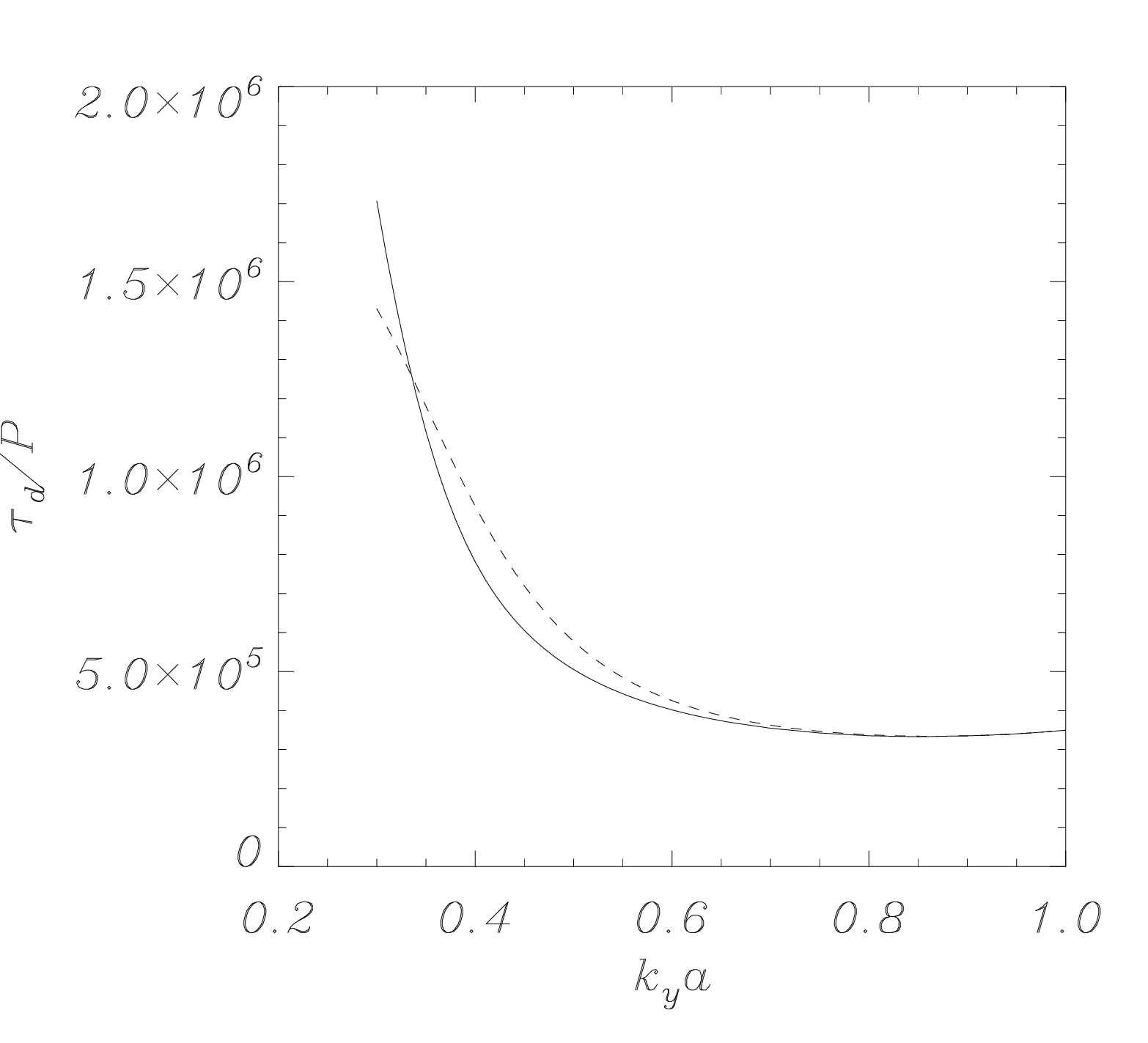}
\caption{{\em Left}, Imaginary part of the frequency as a function of the position of the slabs centers
for the kink solutions ({\em top}), the sausage surface solutions 
({\em middle}), and the sausage body solutions ({\em bottom}) 
for  $k_ya=0.3$ and $l/a=0.3$. 
{\em Right}, Damping time over period as a
function of the perpendicular wavenumber 
for the same six types of solution, for the same parameter values and a fixed distance between 
the slabs, $x_0=5a$. In all the figures the solid lines correspond to the symmetric solution and the 
dashed lines to the antisymmetric solution. A value for the magnetic Reynolds number $R_m=v_{Ai}a/\eta=10^{7}$ has 
been used in a non-uniform grid with $N_x=14\ 000$ points ($2\ 000$ in each of the four non-uniform transitional layers) 
in the range $-80\leq x/a\leq80$.}
\label{dampingw}
\end{figure}

We now consider the six different solutions presented in \S~\ref{undamped} and compute their damping due 
to the resonant coupling to Alfv\'enic motions. The same parameter values for the equilibrium configurations 
are considered, but now non-uniform transitional layers of thickness $l/a=0.3$ have been included.
The results of our computations are displayed in Figure~\ref{dampingw}. The left panels of this figure show
the imaginary part of the frequency, for the six collective solutions under consideration, as a function of the position 
of the centers of the slabs. As can be seen, the splitting of solutions into symmetric and antisymmetric modes produces
these modes to have a damping different from that of the corresponding single slab solutions.
These differences go to zero in the limit of large separation between the slabs. As for the kink solutions, the 
antisymmetric mode has a stronger damping than its symmetric counterpart. This is also true for the sausage-like
solutions. The most significant differences on the damping of the kink symmetric solution are produced for
small distances between the slabs. However, the difference of the symmetric mode damping  with respect to the kink mode 
damping of a single slab is below 10\%. The kink antisymmetric solution damping, on the other hand, is more than double
that of the kink mode of a single slab. Also significant are the differences between the sausage surface
symmetric and antisymmetric damping and the corresponding sausage surface mode damping of a single slab. 
The imaginary parts for these two modes are 70\% smaller and larger, respectively, than the value for the single slab 
mode. 
The right panels of Figure~\ref{dampingw} display the damping per period of the six solutions as a 
function of the perpendicular wavenumber, for a fixed value of the distance between the slabs.
As we increase the value of perpendicular
wavenumber the damping for  both symmetric and antisymmetric solutions 
approaches the damping of the kink, sausage surface, and sausage body solutions of a 
single slab, in the limit of quasi-perpendicular propagation. 
Apart from the difference in the damping of these solutions with respect to those of the single slab and the 
dependence of the damping with perpendicular
wavenumber, the most 
important result is that surface-like solutions, kink and sausage alike, are clearly strongly damped, 
within a few periods, for large values of $k_y$, while sausage body
solutions are  unaffected by resonant couplings (see Figure~\ref{dampingw}, bottom row). The reason for the absence
of resonant damping in the case of the two sausage body solutions is that their frequencies lie outside the 
Alfv\'en continua produced by the presence of the four non-uniform layers. This leads to damping times, for these sausage 
solutions, that are of the order of the used Reynolds number and several orders of magnitude larger that the periods 
of oscillation.

\begin{figure}
 \includegraphics[width=8.0cm,angle=0]{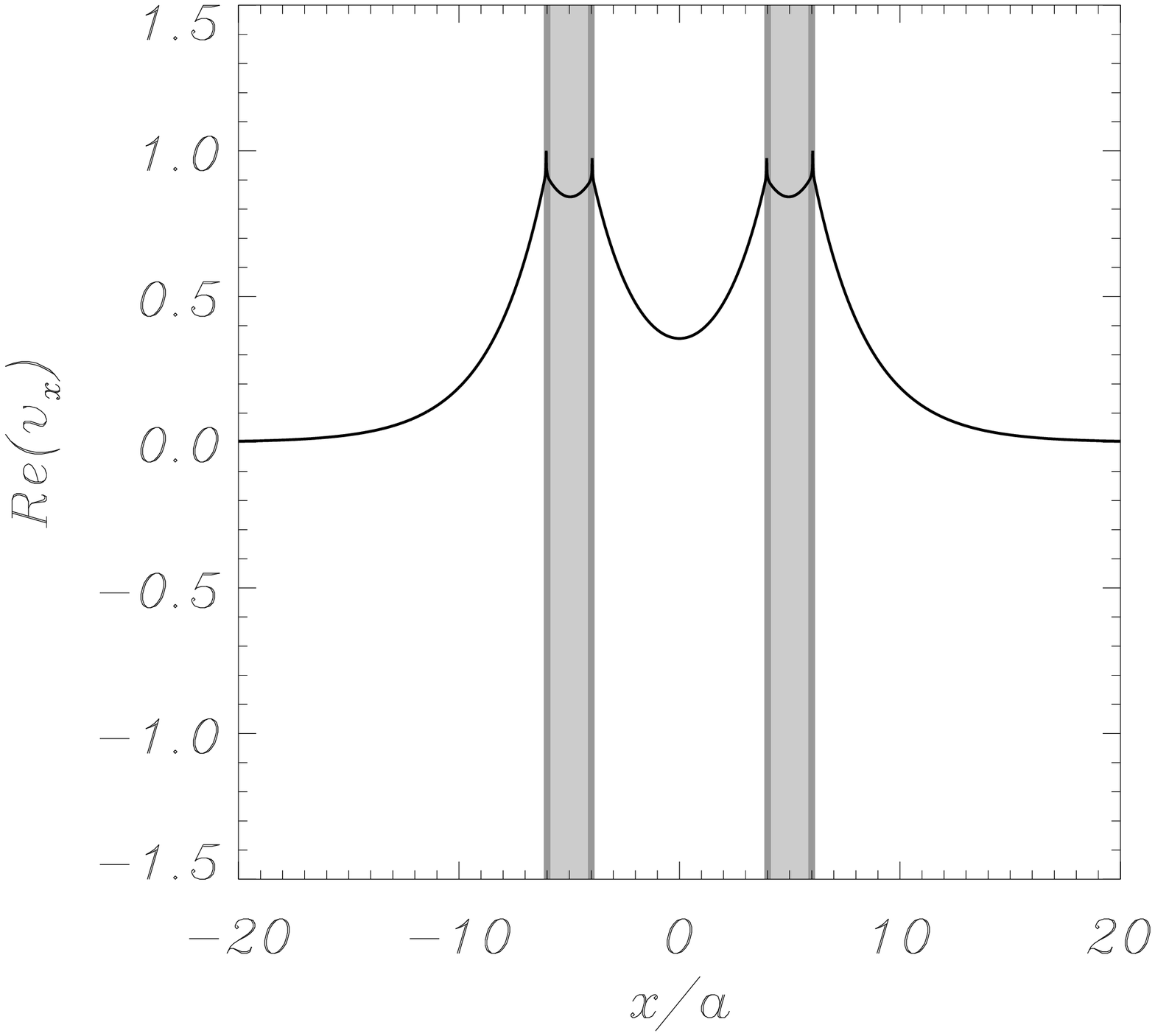}
\includegraphics[width=8.0cm,angle=0]{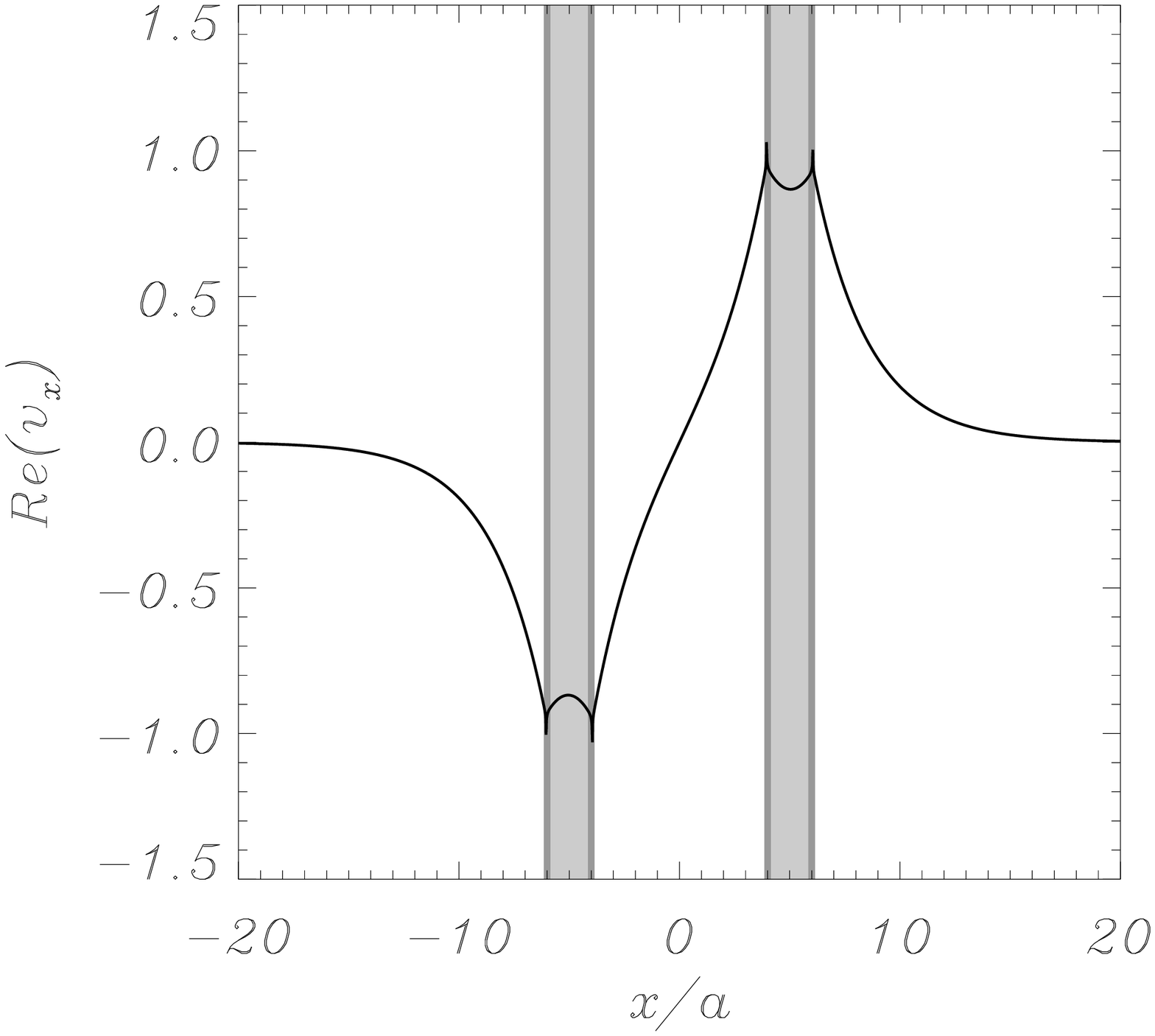}\\
\includegraphics[width=8.0cm,angle=0]{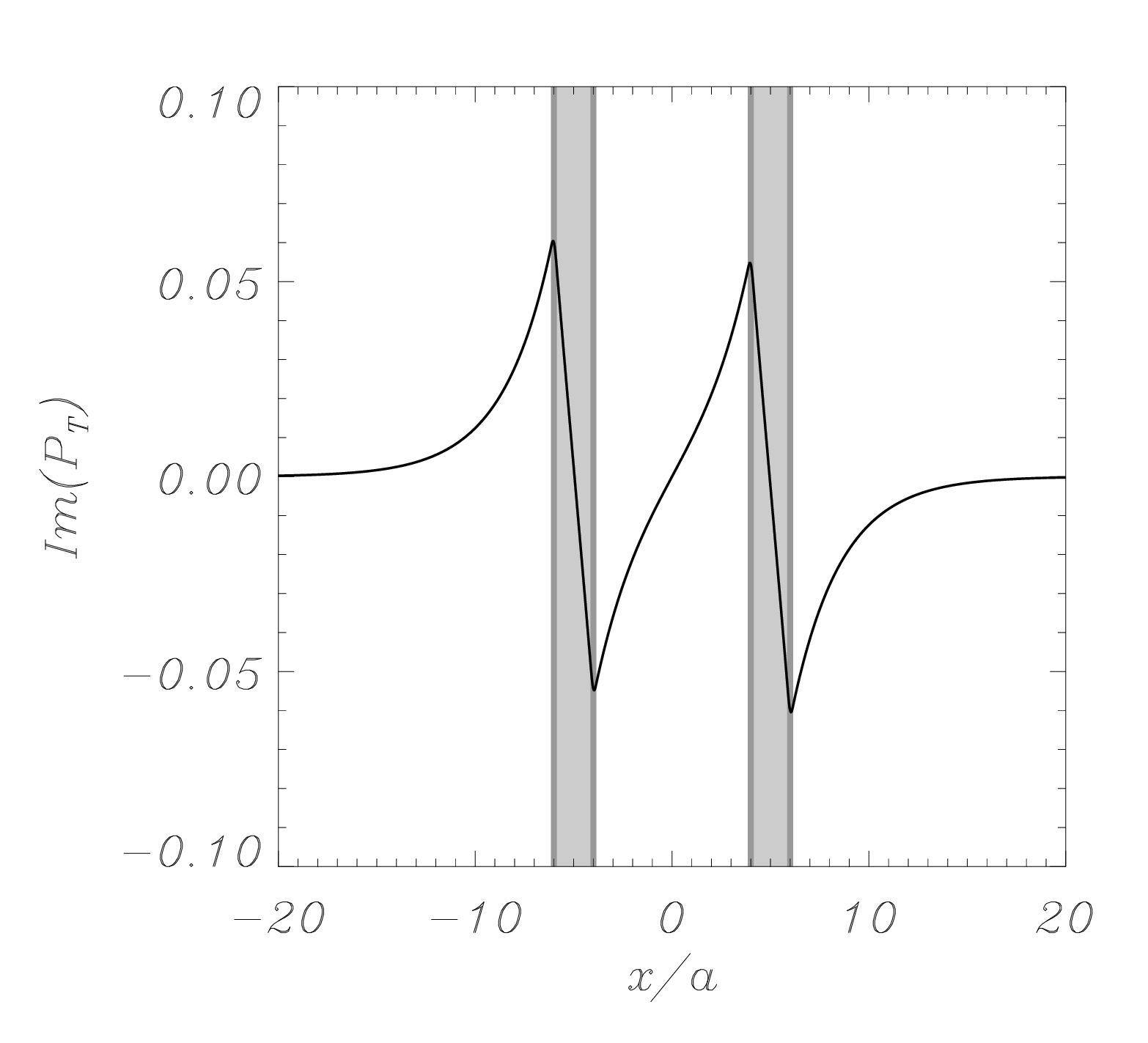}
\includegraphics[width=8.0cm,angle=0]{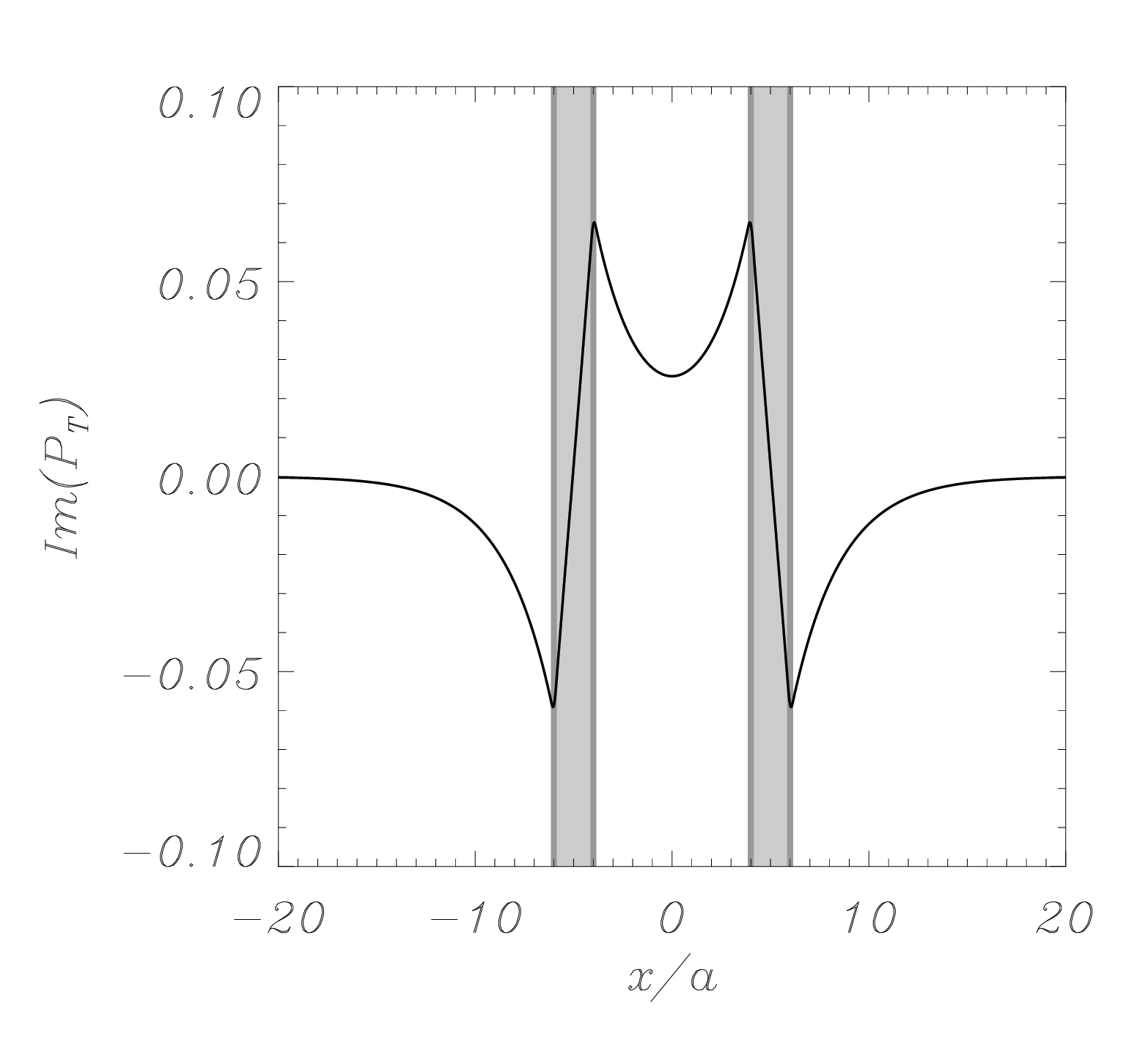}
\caption{Real part of the transversal velocity component, $v_x$, and imaginary part of the 
perturbed total pressure, $P_T$, for the fundamental symmetric ({\em left}) and antisymmetric 
({\em right}) kink solutions, for $k_ya=0.3$ and non-uniform transitional layers with thickness $l/a=0.3$.
 A value for the magnetic Reynolds number $R_m=v_{Ai}a/\eta=10^{7}$ has been
used in a non-uniform grid with $N_x=14\ 000$ points ($2\ 000$ on each of the four non-uniform transitional layers) in the range $-80\leq x/a\leq80$.
In all figures the light-shaded regions represent the density enhancements and the shaded regions the non-uniform layers.}
\label{eigenkinkkinkdamp}
\end{figure}
\begin{figure}
 \includegraphics[width=8.0cm,angle=0]{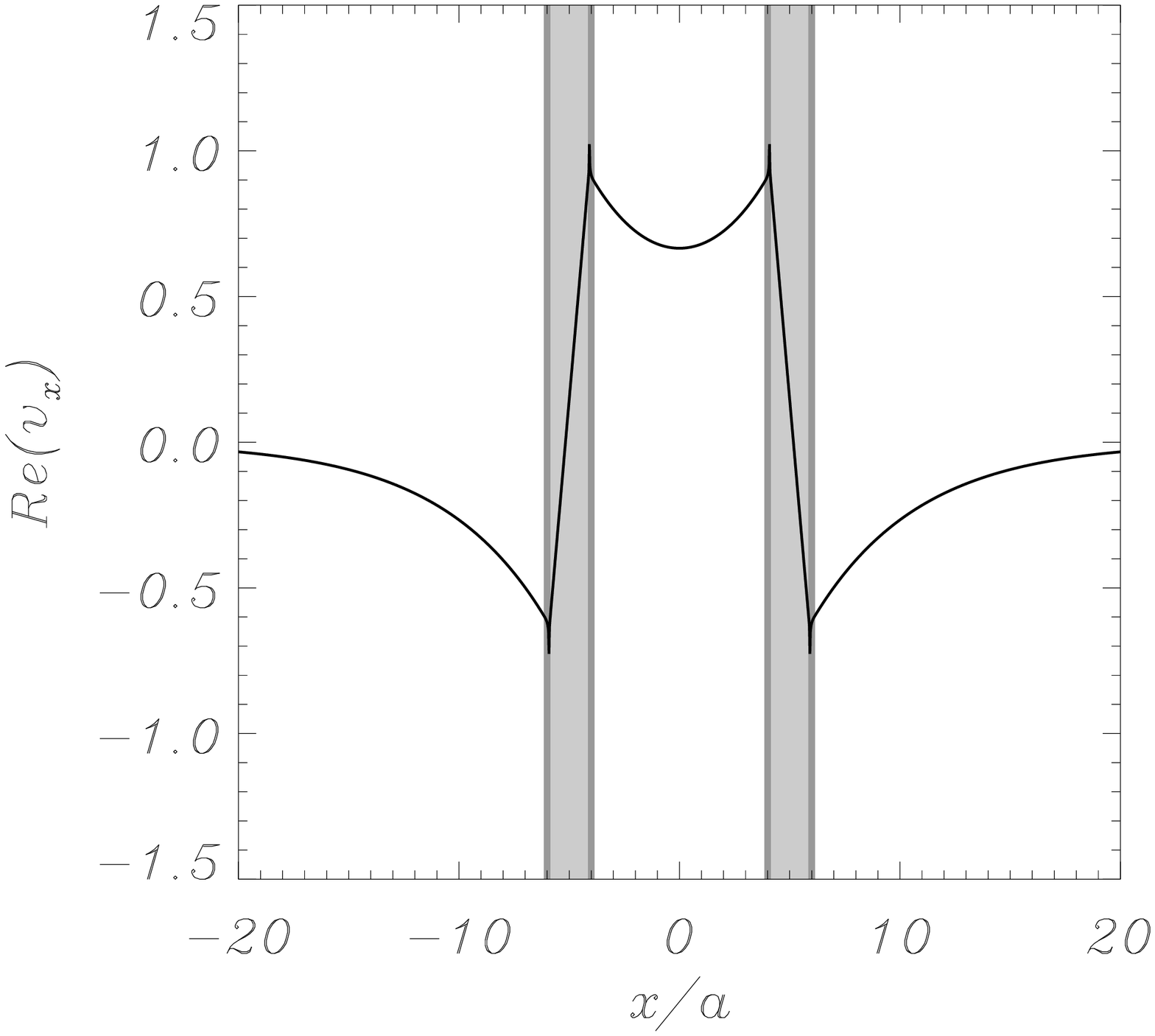}
\includegraphics[width=8.0cm,angle=0]{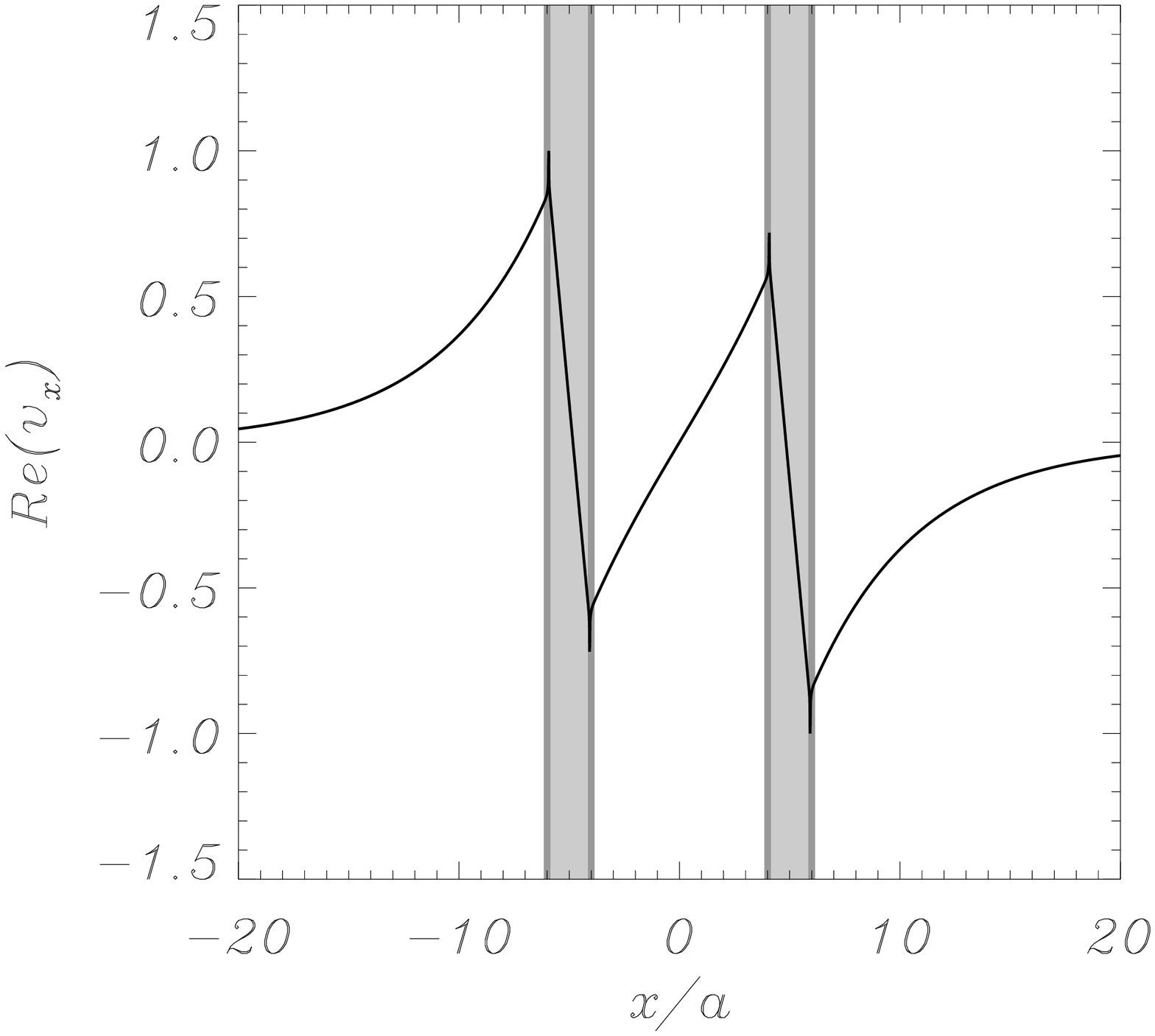}\\
\includegraphics[width=8.0cm,angle=0]{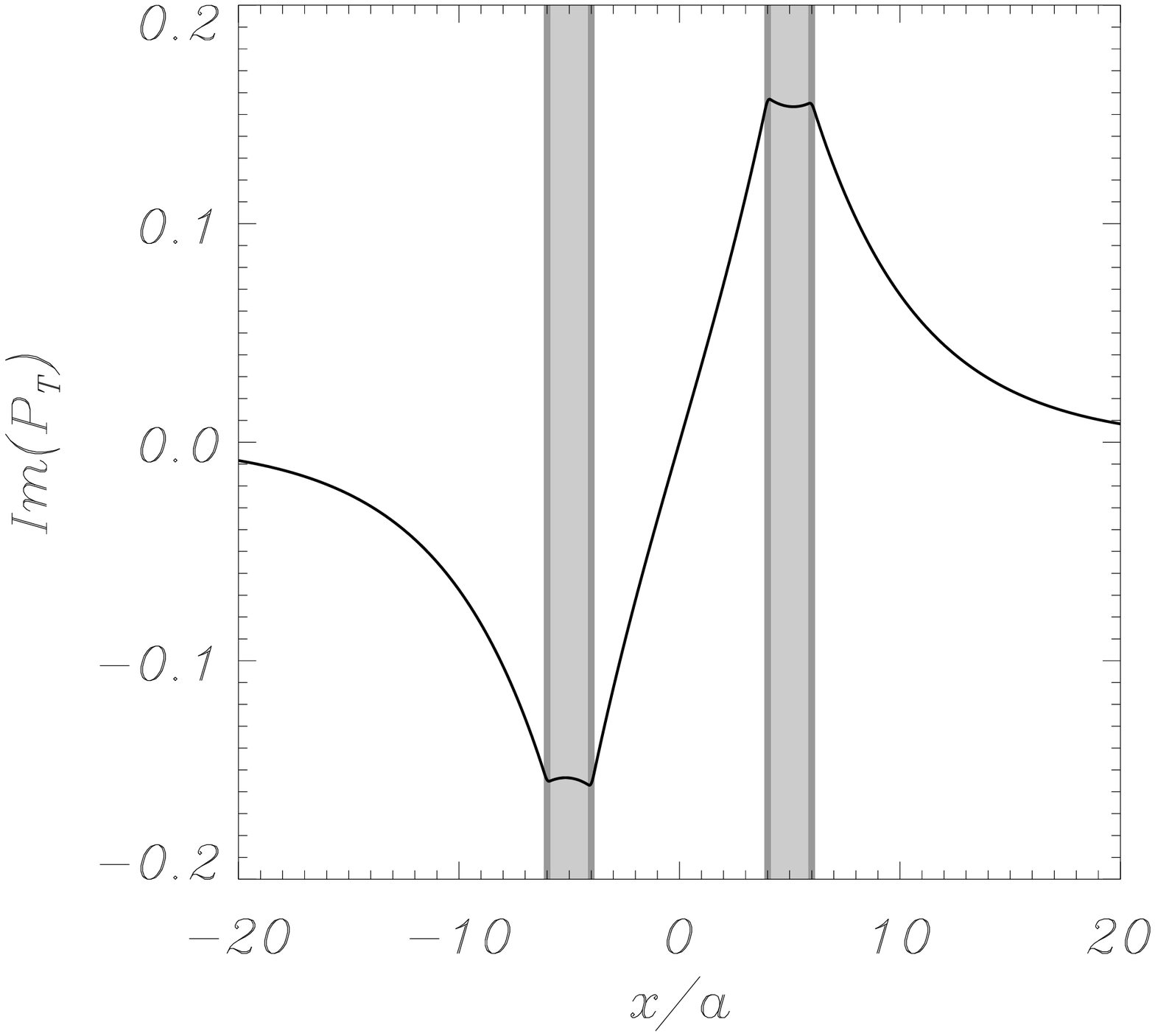}
\includegraphics[width=8.0cm,angle=0]{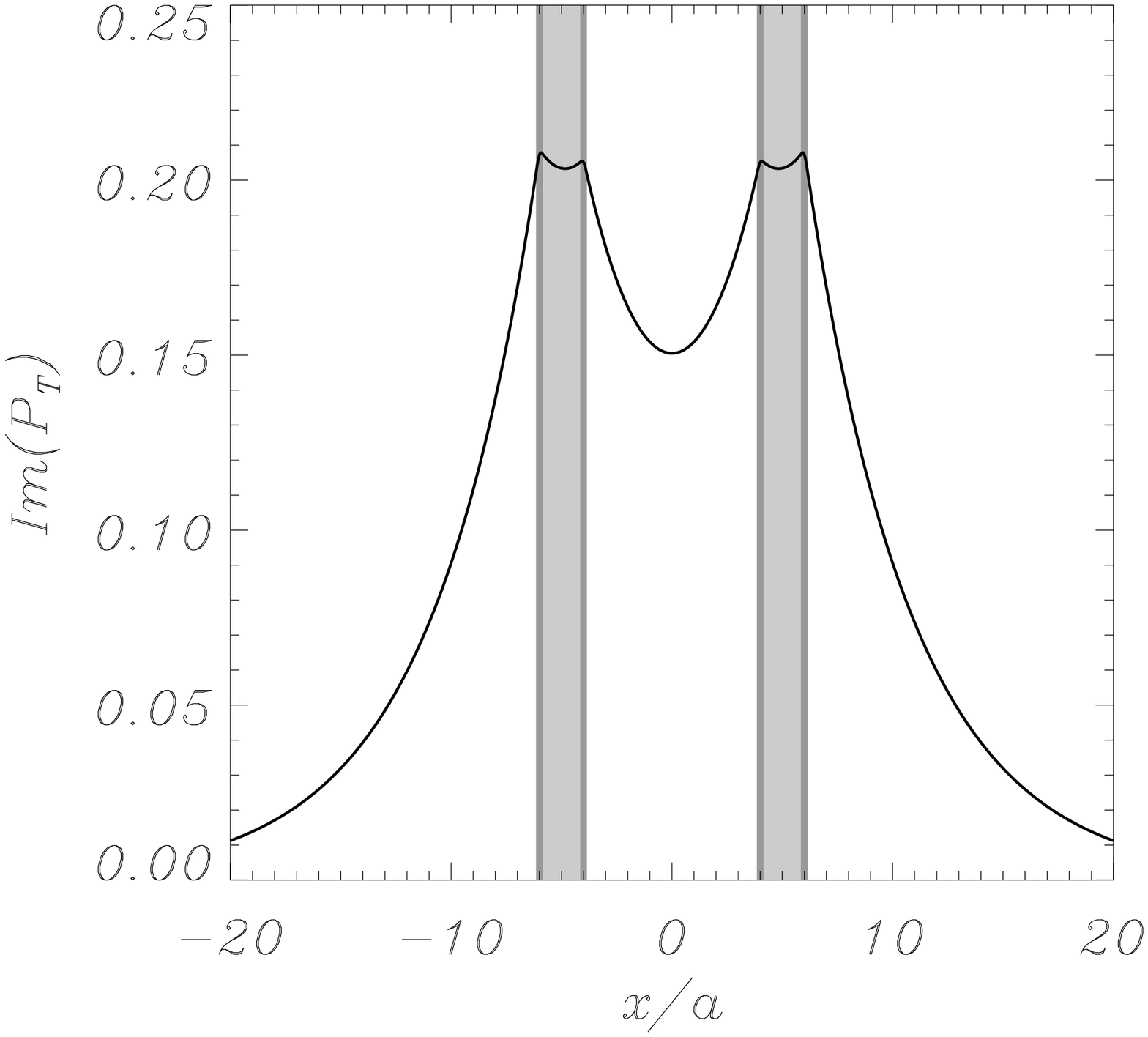}
\caption{Same as Figure~\ref{eigenkinkkinkdamp} for the fundamental symmetric ({\em left}) and antisymmetric ({\em right}) 
sausage surface solutions.}
\label{eigensusudamp}
\end{figure}
\begin{figure}
 \includegraphics[width=8.0cm,angle=0]{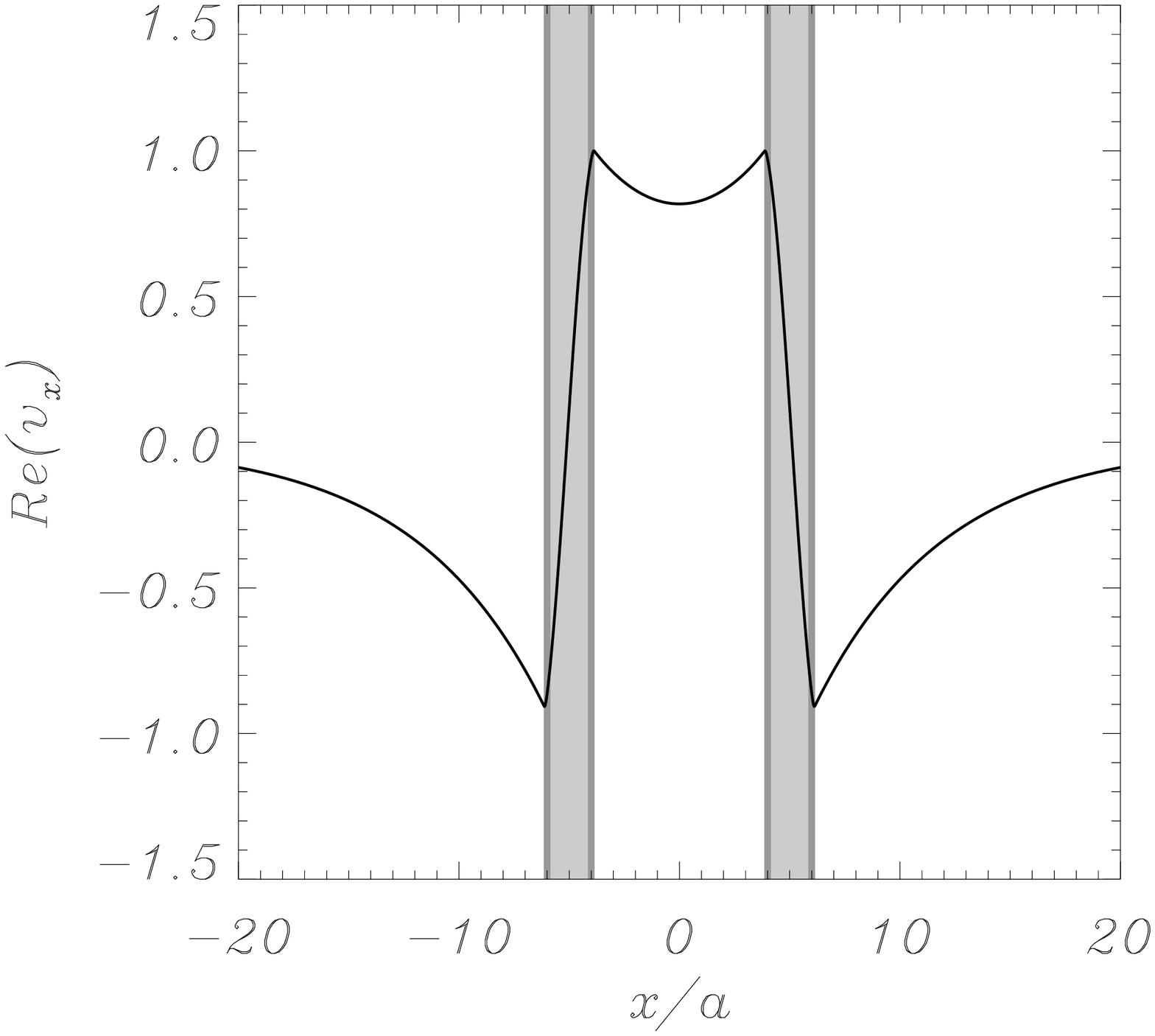}
\includegraphics[width=8.0cm,angle=0]{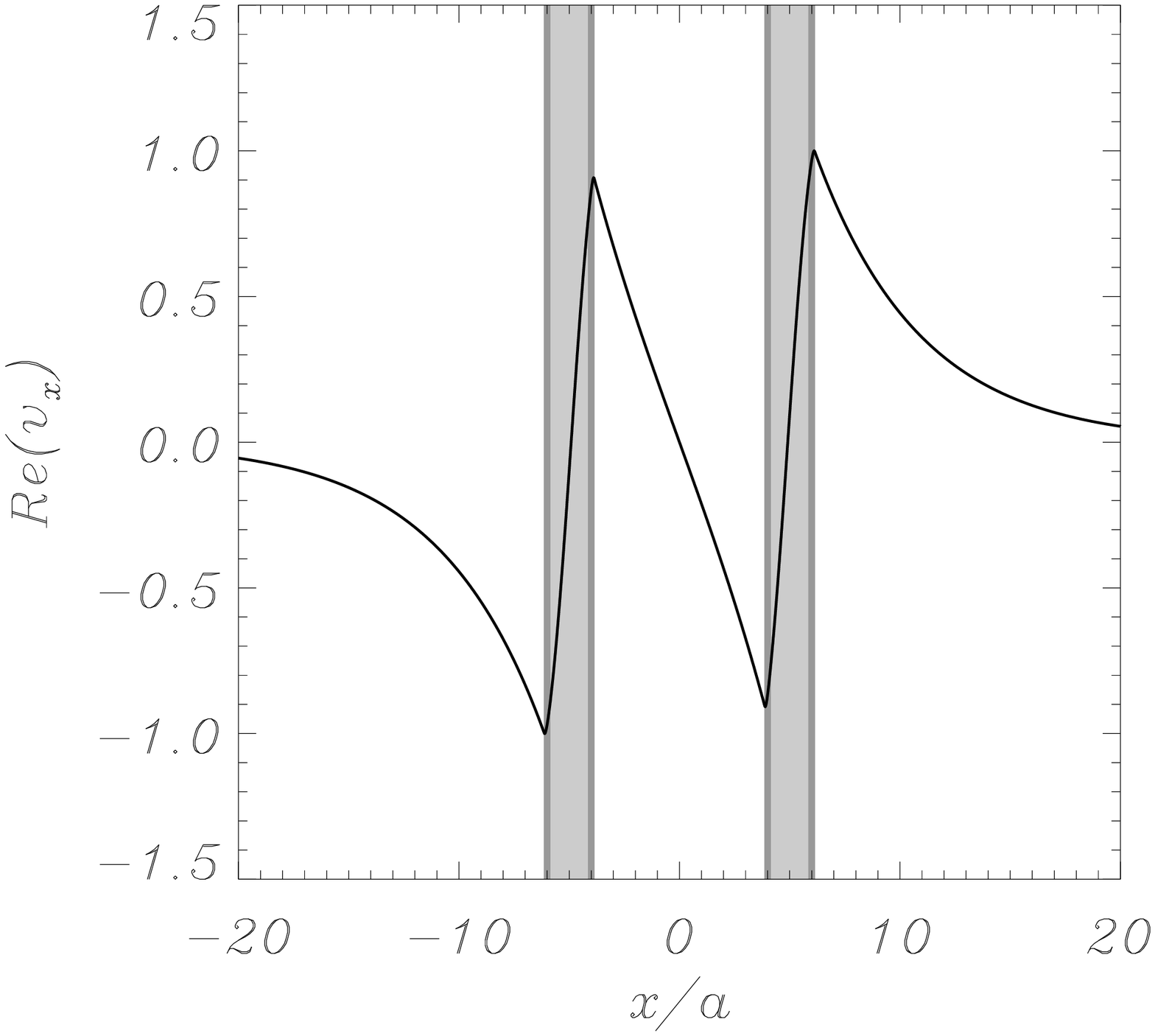}\\
\includegraphics[width=8.0cm,angle=0]{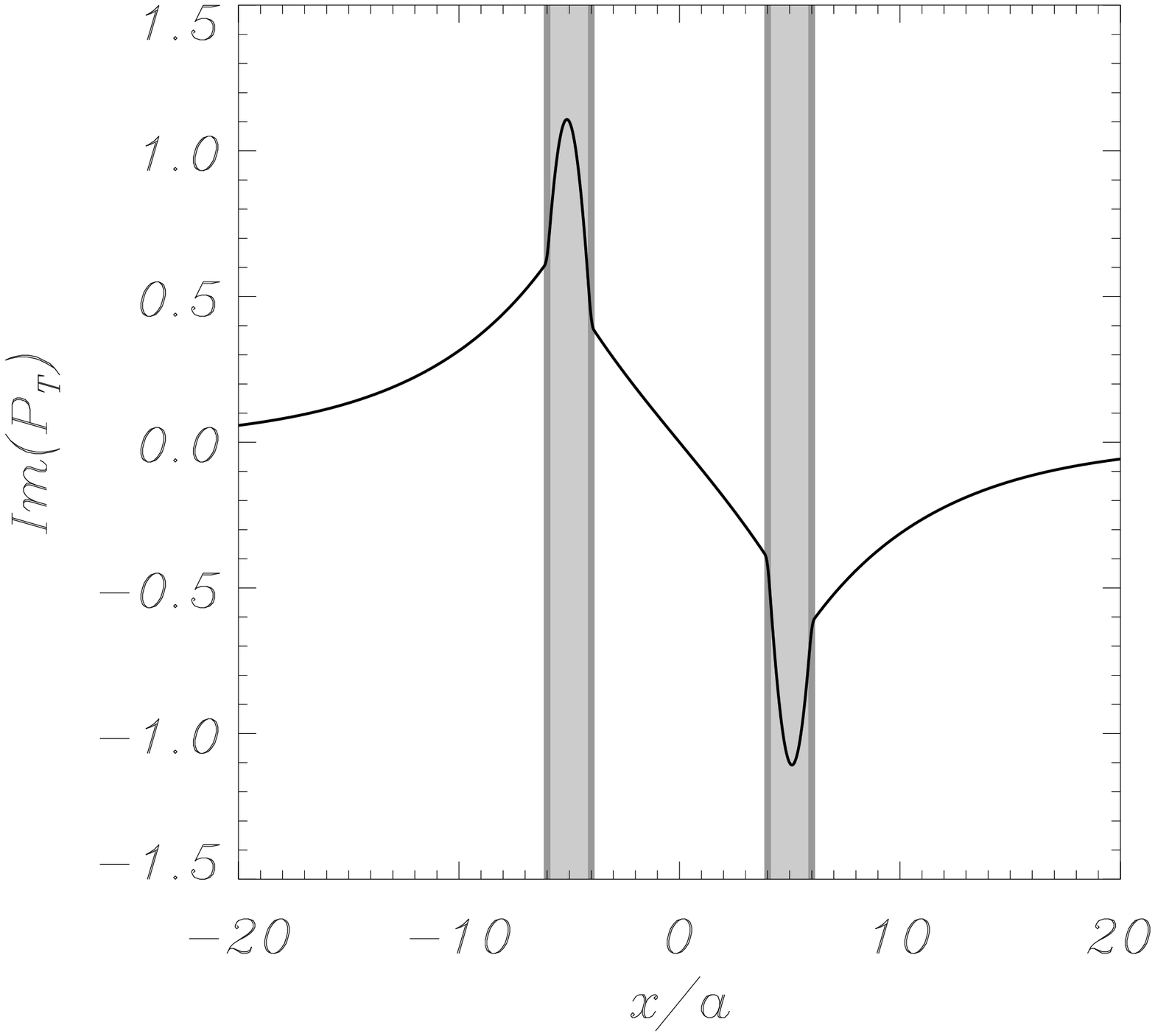}
\includegraphics[width=8.0cm,angle=0]{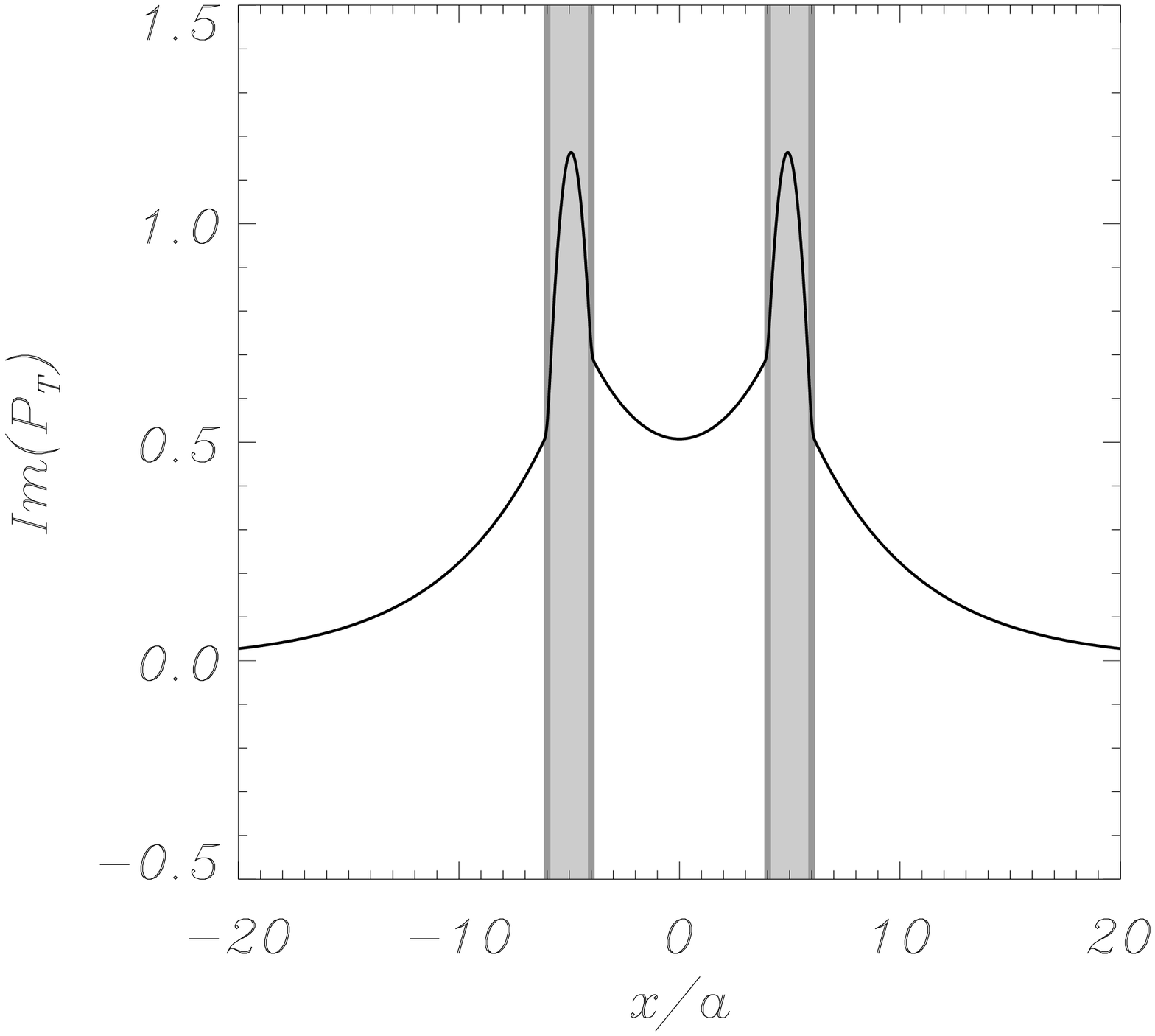}
\caption{Same as Figures~\ref{eigenkinkkinkdamp} and \ref{eigensusudamp} for the fundamental symmetric ({\em left}) and antisymmetric 
({\em right}) sausage body solutions.}
\label{eigensaussausdamp}
\end{figure}

In order to better understand the results displayed in Figure~\ref{dampingw} and to get physical insight into the 
reason(s) for the different damping rates of the different solutions, we have analyzed 
the spatial distribution of the eigenfunctions for the six kind of solutions. 
Figures~\ref{eigenkinkkinkdamp}, \ref{eigensusudamp}, and \ref{eigensaussausdamp} display example
eigenfunctions for the six types of solutions, for a particular set of parameters. 
The spatial distribution of eigenfunctions is rather similar to the ones shown in Figures~\ref{kinkkink}, 
\ref{ssss}, and \ref{sbsb}, for undamped solutions, but 
now peaks at the non-uniform transitional layers are clearly visible in the perturbed transverse 
velocity of the kink and sausage surface modes, while they are absent in the case 
of the two sausage body modes. These peaks are an indication of resonant coupling to 
Alfv\'en waves at the non-uniform layers. Aside from these peaks, the spatial distribution of the magnetic 
pressure perturbation has extrema at the four resonant layers in the case of the kink  and 
the sausage surface modes, while it is maximum at the centers of the slabs in the case
of the sausage body modes. The total pressure perturbation has no extrema at the resonant layers in the case of the 
sausage body solutions. The behavior of the eigenfunctions at the resonant layers and the fundamental conservation laws
that govern the resonant couplings were studied by \citet{sakurai91} 
in ideal MHD and by \citet{goossens95} in dissipative MHD. These authors find that the derivative of the total pressure 
perturbation is zero at the resonant positions. This behavior is retrieved in the pressure perturbation 
for our damped solutions, but not in the two solutions for which resonant couplings are absent.

\section{SUMMARY AND CONCLUSIONS}\label{conclusions}


We have presented numerical results from a normal mode analysis for resonantly damped fast MHD 
oscillations in a system of two coronal loops. For simplicity, the adopted 
equilibrium model consists of two identical density enhancements in Cartesian 
geometry. The linearized resistive MHD wave equations are considered and solutions 
for obliquely propagating waves are obtained. The inclusion of oblique propagation together with non-uniform
transitional layers that connect the internal and external densities produce the resonant damping of 
oscillations.


The considered equilibrium supports a variety of different wave solutions. The normal modes
of oscillation in a two-slab system are, in general, different  to those of a single and isolated slab, due to the
interaction between the slabs. The well-known fast kink 
and sausage normal modes, present in a single slab configuration, split into symmetric and 
antisymmetric solutions. We have concentrated our analysis on the six solutions coming from the splitting
of the kink  and sausage (body and surface) modes in a single slab with oblique propagation, described by 
\citet{ATOB07b}.
The frequencies of these six symmetric and antisymmetric solutions are in general different from the single slab 
solutions. They depend upon the distance between the slabs (as shown by \citealt{Luna06}, for the kink 
symmetric  and antisymmetric solutions), but also
on the magnitude of the perpendicular wavenumber.
The inclusion of oblique propagation of perturbations has important effects on the 
dynamics of the system. First, it introduces two new surface-like solutions, with sausage parity,  
not present in the case of parallel propagation. Second, it decreases the effect of the interaction 
between the slabs, since solutions are more confined to the neighborhood of the density enhancements.  
Both the distance between slabs and perpendicular propagation determine the difference in frequency between the 
single slab model solutions and the modes described in this paper. These differences are of the order of 15\% at most 
and should, therefore, be taken into account in the modeling of these events. They are, however,  well below the 
observational uncertainties for the measured periods in 
transversal coronal loop oscillations, which are of the order of 40\% \citep{Aschwanden02}.  


Next, we have studied the damping of the different solutions due to resonant coupling to Alfv\'enic 
motions. The period and damping rate of the six solutions have been computed and the spatial structure of 
eigenfunctions analyzed in order to find some physical insight into the obtained results. We find that surface-like 
solutions, both kink and sausage, are strongly damped. On the other hand, the two 
sausage body solutions are unaffected by resonant absorption. 
This is due to the absence of resonance due to the fact that the frequencies of these solutions are outside the corresponding Alfv\'en continua.
A comparison with the single slab model results reveals that differences in damping are important for small values of 
the distance between the slabs. The most significant differences arise for the kink antisymmetric solution and 
the two sausage surface solutions, with differences in the imaginary part of the frequency of the order 
of 125\% and 70\% with respect to the damping of the corresponding single slab modes. 
These differences are of the order or larger than the observational uncertainties for the measured damping times in 
transversal coronal loop oscillations, which are around 60\% \citep{Aschwanden02}.




In this work a magnetic two-slab system has been used to model a two-loop structure.
Some of the results found in slab geometry cannot be translated to cylindrical coronal loop systems, 
while some others give a good approximation to the phase speeds in cylinders. For instance, the 
sausage surface solutions described in this paper are not present in a system of two 
cylinders. Also, the sausage body solutions would correspond to perturbations in each 
cylinder with $m=0$, thus they cannot be resonantly damped solutions. On the other hand, slabs are known
to be poor wave-guides in comparison with cylinders. However, our results indicate that the inclusion 
of oblique propagation produces a sharper drop-off rate of the  eigenfunctions in the 
external medium, resulting in a confinement similar to the case of cylinders and a good approximation of
the phase-speed of the kink mode in a slender flux tube. The normal modes of oscillation in a system of two
homogeneous cylinders have been computed recently by \citet{Luna07}. These authors find four solutions
that correspond to transversal oscillations, symmetric and antisymmetric, in the plane containing the two 
cylinders. The frequency of these solutions is seen also to differ from the frequency of the kink mode
for a single cylinder for small distances between the cylinders. A natural extension to the study 
presented in this paper and to the results described by \citet{Luna07} is the analysis of the damping 
in a similar two-cylinder configuration.

\acknowledgments

The authors acknowledge the Spanish Ministerio de 
Educaci\'on y Ciencia for the funding 
provided under project AYA2006-07637 and the Conselleria d'Economia, Hisenda i 
Innovaci\'o of the Government of the Balearic Islands for the funding provided under 
grants PRIB-2004-10145 and PCTIB2005GC3-03 (Grups
Competitius). J. Terradas acknowledges the Spanish Ministerio de 
Educaci\'on y Ciencia for the funding provided under a Juan de la Cierva fellowship.


\clearpage

\clearpage

\clearpage

\clearpage

\clearpage

\clearpage

\clearpage

\clearpage

\clearpage

\end{document}